\documentclass[aps,pra,twocolumn,showpacs]{revtex4}

\newcommand{\ket}[1]{| \, #1 \rangle}

\newcommand{\bra}[1]{\langle #1 \, |}

\def\>{\rangle}
\def\<{\langle}

\newcommand{\be}{\begin{equation}}
\newcommand{\ee}{\end{equation}}
\newcommand{\bea}{\begin{eqnarray}}
\newcommand{\eea}{\end{eqnarray}}
\newcommand{\gr}[1]{\boldsymbol{#1}}
\renewcommand{\vec}[1]{{\gr{#1}}}

\usepackage{graphicx,subfigure}
\usepackage{amsmath}
\usepackage{amsfonts,amssymb,graphicx,latexsym,color}
\usepackage{bm,bbm}

\graphicspath{{Images/}}

\begin{document}

\title{Modelling of quantum information processing with Ehrenfest guided trajectories:\\ a case study}
\author{Sai-Yun Ye,$^{1}$ Dmitrii Shalashilin,$^{2}$ and Alessio Serafini$^{1}$}
\affiliation{$^{1}$ Department of Physics and Astronomy, University College London, Gower Street, London WC1E 6BT, UK}
\affiliation{$^{2}$ School of Chemistry, University of Leeds, Leeds LS2 9JT, UK}


\date{\today }

\begin{abstract}
We apply a numerical method based on multi-configurational Ehrenfest trajectories, and demonstrate
converged results for the Choi fidelity of an entangling quantum gate between two two-level systems interacting
through a set of bosonic modes.
We consider both spin-boson and rotating wave Hamiltonians, for various numbers of mediating modes (from $1$ to $100$),
and extend our treatment to include finite temperatures.
Our results apply to two-level impurities
interacting with the same band of a photonic crystal, or to two distant ions interacting with the same set
of motional degrees of freedom.
\end{abstract}
\maketitle

\section{Introduction}

The ability of tracking the evolution of complex quantum systems will be a crucial
support to the design and development of future quantum technologies.
A paradigm of particular interest for the latter is one where finite dimensional quantum systems,
typically two-level systems (qubits), interact through a `bus',
made up of a set of bosonic field modes \cite{cirac95,cirac97,vanenk,1,martin99,13,zheng00,17,wineland,
clark,bla,xiao,mauro04,barrett,20,zhou,beige1,serafozzi,qubus,27,6,zou,28,beige2,26,
chalapat,8,zhenbiao,noiartri,noiartri2,hybrid,crystal,2cav}.
In this paper, we will consider the case of two qubits interacting with a common discrete set of
modes in the non-perturbative regime. While the possibility of creating entanglement, even at steady state,
by the interaction with a common bosonic bath has been highlighted repeatedly in the literature,
the case of non-perturbative interactions with a bunch of $10$ or $20$ modes
(as will be the case in our study) presents several major technical difficulties, essentially due to the
impossibility of an analytic master equation approach -- only possible in the continuum limit under the
Born-Markov approximation -- and to the huge size of the dynamically relevant part of the Hilbert space.

Various numerical approaches have been developed to emulate these dynamics on classical computers,
such as the so called multiconfigurational time dependent Hartree method \cite{mctdh,wang00,wang08} and its ``Gaussian'' variation \cite{gmctdh},
various schemes based on path integral techniques \cite{makri98,nesi07}, and even the adaptation of time-adaptive density matrix renormalisation group
techniques \cite{prior10} borrowed from condensed matter theory.
Here, we will tackle such difficulties by
borrowing a method co-pioneered and developed by one of the authors in the arena of
chemical physics \cite{shalashilin09,shalashilin10}. The method is based on the adoption of a
set of tensor products of time-dependent coherent states as a discrete `basis grid' on which to represent the field degrees of freedom (referred to as ``coupled coherent states''
in the literature \cite{ccs}), and on letting the states of the co-moving grid evolve according to their Ehrenfest dynamics
(whose application to a grid of coherent states goes under the name of ``multi-configurational
Ehrenfest'' method). This approach has the advantage of being relatively light in terms of
computational resources, easy to program, and yet
of allowing one to follow coherent quantum dynamics in detail, as it will be shown.

In our study, we will focus on a specific, but very relevant,
aspect of the quantum dynamics of the two qubits: we shall consider the realisation of
an entangling quantum gate between them, namely of a controlled Z (CZ) gate.
To estimate the quality of such a realisation we will consider the quantum fidelity between
the pure state corresponding to the CZ gate by the standard channel-state
duality (Choi isomorphism \cite{matt,choi75,belavkin86}) and the quantum state corresponding to the
channel acting on the two qubits upon partial tracing over the field's degrees of freedom.

Our main aim is demonstrating the capability of Ehrenfest guided trajectories in phase space
to produce reliable and converged results for complex figures of merit, able to reveal detailed information about the quantum dynamics of the constituents.
The `Choi fidelity' of a two-qubit quantum gate is a property of the dynamics itself, and not of the initial state, and its evaluation requires, at any time, the evolution of ten initial states: it is, therefore, a rather cumbersome figure of merit to compute, let alone to optimise over a rather large range of values of the dynamical parameters, as we did.
The advantages of the Ehrenfest guided trajectories over -- arguably more precise but heavier -- approaches based on full variational principles are manifest in such circumstances.

As for direct impact, let us remark that our study would apply on systems like
two-level impurities trapped in a photonics crystal and interacting with the same band of allowed modes \cite{natcrystals,crystal,crystals}, or
to the internal levels of two ions interacting with all the vibrational modes of an array of ions in a linear trap \cite{spinions}. It is important to point out that our treatment can account for finite, although relatively small, temperatures as well.

Our paper is organised as follows. In section \ref{ehrenfest} we will review the basic theory behind methods
of solution of the Schr\"odinger equation based on a set of time-dependent, Ehrenfest guided basis states.
We will not dwell so much on the technical details, which are covered elsewhere, as on the basic concepts,
and will try to present them in terms which will be friendly to an audience with no previous familiarity with the terminology of chemical physics or molecular dynamics.
In section \ref{model} we will introduce the physical Hamiltonian and define precisely our chosen figure of merit.
Section \ref{results} will contain the main results of our numerical study. Finally, we will draw conclusions, and discuss advantages and shortcomings of our method of choice, in
section \ref{conc}.

\section{Ehrenfest guided trajectories \label{ehrenfest}}

The main difficulty in dealing with a system including
few two-level systems and a bunch of $M$ bosonic modes is clearly how to
handle the infinite dimensional bosonic Hilbert space.
The method we apply here, referred to in the literature as  `multi-configurational
Ehrenfest' and first introduced in \cite{shalashilin09}, tackles this difficulty on the shoulders of two major assumptions:
\begin{itemize}
\item[i)] the state space of the field modes is represented as a superposition of $N$ time dependent coherent states;
\item[ii)] the time dependence of the coherent states is determined by
a simplified variational principle (which, in other words,
dictates how the finite dimensional subspace spanned by the set of coherent states changes with time,
trying to keep it in the dynamically relevant region).
\end{itemize}
The finite dimensional systems involved are, on the other hand, treated by representing their entire Hilbert space,
spanned by $d$ orthonormal basis states $\ket{l}$,
for $l\in\{1,d\}$.
In agreement with i), the ansatz wave-function of the whole system reads
\be
\ket{\psi} = \sum_{l=1}^{d} \sum_{j=1}^{N} c_{l,j}(t) \left(\ket{l}\otimes\ket{\vec{\alpha}_j(t)}\right) \; ,
\ee
where $\vec{\alpha}_j\in{\mathbbm C}^M$ $\forall\, j$ and  each $\ket{\vec{\alpha}_j}$ is a tensor product of
coherent states: $\ket{\vec{\alpha}_j} = \bigotimes_{m=1}^{M} \ket{\alpha_j^{(m)}}$, such that,
if $a_m$ is the annihilation operator of mode $m$, one has $a_{m}\ket{\vec{\alpha}_j}=\alpha_j^{(m)}\ket{\vec{\alpha}_j}$. Since we will be dealing with two qubits, it will be $d=4$ for us.

The evolution of the dynamical parameters is more conveniently described by adopting a Lagrangian formulation.
For a Hamiltonian operator $\hat{H}$, let us define ${\mathcal L}$ as
\be
{\mathcal L} = \bra{\psi}\hat{H}-i\partial_t\ket{\psi} \; . \label{fullag}
\ee
The time-derivative operator is defined as the differentiation of the time dependents coefficients $c_{l,j}$
and by the relationships $\partial_t \ket{l}=0$ (the finite dimensional system's basis is time-independent)
and $\partial_t \ket{\vec{\alpha}_j}=\sum_{m=1}^{M} [\dot{\alpha}_j^{(m)}(a_m^\dag-\alpha_j^{(m)*}/2)-
\dot{\alpha}_j^{(m)*}\alpha_j^{(m)}/2]\ket{\vec{\alpha}_j}$ (derived from the time-dependence of a coherent state with varying
phase space position).
The quantity ${\mathcal L}$ is hence a function of the coefficients $c_{l,j}$, the complex parameters
$\vec{\alpha}_j$ and their time-derivatives $\dot{c}_{l,j}$ and
$\dot{\vec{\alpha}}_j$. In fact, it can be shown that ${\mathcal L}$ serves as a Lagrangian
for the quantum system, in the sense that the Euler-Langrange equations
\be
\frac{\partial{\mathcal L}}{\partial c_{l,j}} = \frac{\rm d}{{\rm d}t} \frac{\partial \mathcal L}{\partial \dot{c}_{l,j}}
\ee
are {\em equivalent} to Schr\"odinger equation \cite{lastfootnote}. See Appendix \ref{schroed} for more details.

Besides determining the state evolution, the variational formalism also provides one with a recipe to update the basis such that
the expression ${\mathcal L}$ of Eq.~(\ref{fullag}), which clearly always equals $0$ in the exact dynamics, is minimised during the
time-evolution. Such a minimisation, which in essence keeps the basis in the `most relevant' region of the Hilbert space within the constraints of the adopted approximation, would be obtained by considering the full Euler-Lagrange equations for the $M\times N$ complex parameters
$\vec{\alpha}_j$ and their time-derivatives. This would be a large nonlinear system of coupled equations, requiring a substantial numerical effort to be solved. Instead, we introduce here assumption ii), and replace the full variational equations for $\vec{\alpha}_j$ with
a simplified version thereof. In particular, we will neglect all terms coupling the different $\vec{\alpha}_j$'s, on the grounds that
the overlaps $\bra{\vec{\alpha}_j}\vec{\alpha}_k\rangle$ are typically very small if the number of modes $M$ is large enough.
For each $j$, let us then define the vector $\ket{\tilde{\psi}_j}$ as
\be
\ket{\tilde{\psi}_j} = \sum_{l=1}^{d} c_{l,j} \ket{l}\otimes \ket{\vec{\alpha}_j} \; ,
\ee
and the corresponding `approximated' Lagrangian $\tilde{\mathcal L}_j$ as
\be
\tilde{\mathcal L}_j = \bra{\tilde{\psi}_j}H-i\partial_t\ket{\tilde{\psi}_j} \; .
\ee
The equation of motion for the parameter $\alpha_{j}^{(m)}$ (the $m$-th component of the vector $\vec{\alpha}_{j}$) is
\be
\frac{\partial\tilde{{\mathcal L}}_j}{\partial \alpha_{j}^{(m)}} = \frac{\rm d}{{\rm d}t} \frac{\partial \tilde{{\mathcal L}}_j}{\partial \dot{\alpha}_{j}^{(m)}} \; , \label{simpleton}
\ee
where we also neglect the time-dependence of the coefficients $c_{l,j}$, such that
each Lagrangian $\tilde{{\mathcal L}}_j$ only depends on the four complex parameters $c_{l,j}$
and on the $M$ complex parameters represented by the entries of $\vec{\alpha}_j$ (and on their time-derivatives
$\dot{\vec{\alpha}}_j$, see Appendix \ref{ehren} for further details).
Eq.~(\ref{simpleton}) defines the ``Multi-Configurational Ehrenfest'' (MCE) method we are using.

Notice that the assumption ii) is not, per se, an approximation, but rather just a way of choosing the time-dependence of the adopted basis. However, it should be stressed that, in general,
the exact Euler Lagrange equation for the full variation of the parameters $\alpha_j^{(m)}$ is likely to
provide one with a
more accurate result in that it will yield a smaller Lagrangian ${\mathcal L}$ (which is zero in the exact dynamics).

However, Eq.~(\ref{simpleton}) is much easier to treat numerically, hence the advantage of our method, which can be easily programmed and applied with
modest computational resources and often provides results in very good agreement with complete variational methods, like
MCTDH or G-MCTDH \cite{mctdh,wang00,wang08,gmctdh}.

Multi-configurational Ehrenfest guided trajectories have been thoroughly tested for spin-boson dynamics under
different spectral densities, establishing the reliability of their converged results in several, diverse situations \cite{shalashilin09,noiartrisb}.
Here, we will instead apply them to study a composite system including discrete sets of bosonic field modes, where coherent quantum information processing can be carried out.

\section{Model and figure of merit \label{model}}

We set out to study coherent quantum information processing for a system comprising two two-level systems
(``qubits'')
connected by $M$ bosonic modes through a spin-boson like coupling.

In principle, this represents the archetype of a quantum system where complex dynamics and
information processing tasks can be carried out, and whose dynamics is impervious to non-approximated methods.
In practice, our case study may be thought of as representing two two-level atoms (or impurities) interacting
with the same photonic band of a photonic crystal \cite{crystals}, or a simulation of the same setting in a linear array of
trapped ions (where the qubits are embodied by internal levels of the ions interacting with the same
set of vibrational normal modes \cite{spinions}).

For future convenience, let us re-label the four states of the computational basis of the two two-level systems as follows:
\bea
\ket{1} &=& \ket{\downarrow\downarrow} \; , \label{alpha} \\
\ket{2} &=& \ket{\downarrow\uparrow} \; , \label{beta} \\
\ket{3} &=& \ket{\uparrow\downarrow} \; , \label{gamma} \\
\ket{4} &=& \ket{\uparrow\uparrow} \; . \label{delta}
\eea
The operators $\hat{\sigma}_{x}^{(1)}$, $\hat{\sigma}_{x}^{(2)}$, $\hat{\sigma}_{z}^{(1)}$ and $\hat{\sigma}_{z}^{(2)}$,
will stand for the customary Pauli operators in the Hilbert spaces of qubit $1$ and $2$. For instance,
in the adopted basis, $\hat{\sigma}_{x}^{(1)}$ is defined by
$\hat{\sigma}_{x}^{(1)}\ket{1} = \ket{3}$, $\hat{\sigma}_{x}^{(1)}\ket{3} = \ket{1}$,
$\hat{\sigma}_{x}^{(1)}\ket{2} = \ket{4}$, $\hat{\sigma}_{x}^{(1)}\ket{4} = \ket{2}$.

In our study, we shall consider both an actual `spin-boson like' Hamiltonian:
\be
\begin{array}{rl}
\hat{H} =
\sum_{j=1}^{2} \sum_{m=1}^{M} \Big[&\hspace*{-0.1cm}\varepsilon \hat\sigma_{z}^{(j)}+\Delta_j \hat\sigma_{x}^{(j)}
+\omega_{m} a_{m}^{\dag}a_{m}\\
&+\, g^{(j)}_m\sigma_{x}^{(j)}\left(a_{m}+a_{m}^{\dag}\right)\Big] \, , \label{Hsb}
\end{array}
\ee
and its rotating wave counterpart:
\be
\begin{array}{rl}
\hat{H}_{rw} =
\sum_{j=1}^{2} \sum_{m=1}^{M} \Big[&\hspace*{-0.1cm}\varepsilon\hat\sigma_{z}^{(j)}+\Delta_j \hat\sigma_{x}^{(j)} + \omega_{m} a_{m}^{\dag}a_{m} \\
&+ \, g^{(j)}_m\left(\sigma_{+}^{(j)}a_{m}+\sigma_{-}^{(j)}a_{m}^{\dag}\right)\Big] \, , \label{Hrw}
\end{array}
\ee
where $\sigma_{+}^{(j)}=\sigma_{-}^{(j)\dag}=\sigma_{x}^{(j)}+i\sigma_{y}^{(j)}$.
As well known, the Hamiltonian $\hat{H}_{rw}$ is a good approximation of
$\hat{H}$ in the almost resonant, high frequency case that is, in our notation, for
$|2\varepsilon-\omega_m|\ll |2\varepsilon+\omega_m|$, $\forall \, m$.
In the following, we will consider systems with different numbers of bosonic modes $M$,
various values of frequencies $\{\omega_m\}$ and
spin-boson couplings $\{g^{(j)}_m\}$, and different parameters $\Delta_1$ and $\Delta_2$.
Also, we will set $\hbar=1$ throughout the paper.

In reproducing the dynamics of the two qubits
by treating the field through multi-configurational Ehrenfest trajectories, we will aim at obtaining
converged results for a figure of merit of interest in the study of quantum information processing, namely the
fidelity with which an entangling controlled Z (CZ) gate can be realised for the two qubits through the mediating bosonic modes.
In terms of the basis states of Eqs.~(\ref{alpha}-\ref{delta}), a CZ gate is represented as a unitary $U_{CZ}$ leaving all the basis states
invariant except for $\ket{4}$, which becomes $-\ket{4}$, that is
\be
U_{CZ} \ket{j} = f(j) \ket{j} \quad {\rm for} \quad 1\le j \le 4 \, ,
\ee
where $f(j)=1$ for $j\in\{1,2,3\}$ and $f(j)=-1$ for $j=4$.
The relevance of a CZ gate to quantum information processing stems from its being
a maximally entangling gate which, combined with single-qubit unitaries, forms a universal quantum
set for gate based quantum computation \cite{bremner}.

At zero temperature, the quantum operation $\Gamma_t$ we want to compare with the CZ unitary gate is defined as follows,
in terms of a notional initial density matrix of the qubits $\varrho$:
\be
\Gamma_{t}(\varrho) = {\rm Tr}_B \left[ {\rm e}^{-i\hat{H}t} \left(\varrho \otimes \ket{0}\bra{0} \right) {\rm e}^{i\hat{H}t} \right] \, ,
\ee
where ${\rm Tr}_{B}$ stands for partial tracing over the Hilbert space of the bosonic modes and $\ket{0}$ is the vacuum state of the modes.
We will also extend our treatment to include a finite temperature $1/\beta$ of the bosonic modes (in natural units where $k_B=1$),
in which case the quantum operation $\Gamma_{t,\beta}$ will be given by
\be
\Gamma_{t,\beta}(\varrho) = {\rm Tr}_B \left[ {\rm e}^{-i\hat{H}t} \left(\varrho \otimes
\int_{{\mathbbm C}^{2M}}P_\beta(\vec{\alpha})\ket{\vec{\alpha}}\bra{\vec{\alpha}}{\rm d}^{2M}\vec{\alpha} \right) {\rm e}^{i\hat{H}t} \right] \, ,
\ee
with
\be
P_\beta(\vec{\alpha}) = \prod_{m=1}^{M} \left(\frac{{\rm e}^{\beta \omega_m}-1}{\pi} {\rm e}^{-({\rm e}^{\beta\omega_m}-1)|\alpha_{m}|^2}\right) \, . \label{gs}
\ee
The function $P_{\beta}(\vec{\alpha})$ is just the Glauber-Sudarshan P-representation of a thermal state of the bosonic modes (given by the product
of individual P-representation for each of the modes). In our notation $\vec{\alpha}\in{\mathbbm C}^{2M}$, while each component of $\vec{\alpha}$
is denoted by $\alpha_m$.
Clearly, one has that $\lim_{\beta\rightarrow\infty}\Gamma_{t,\beta}=\Gamma_t$.

In our numerical study, we will reproduce the operations $\Gamma_{t}$ by adopting the method detailed in the previous section,
which is defined for pure states, and also reconstruct the operations $\Gamma_{t,\beta}$ by sampling different initial pure coherent states
$\ket{\vec{\alpha}}$ for the field according to the distribution given by $P(\vec{\alpha})$.
We will describe the field in terms of coupled coherent states during the time evolution and then
trace it out to achieve the quantum operation acting on the two qubits.

\begin{figure}[t!]
\begin{center}
\subfigure[]{\includegraphics[scale=0.26]{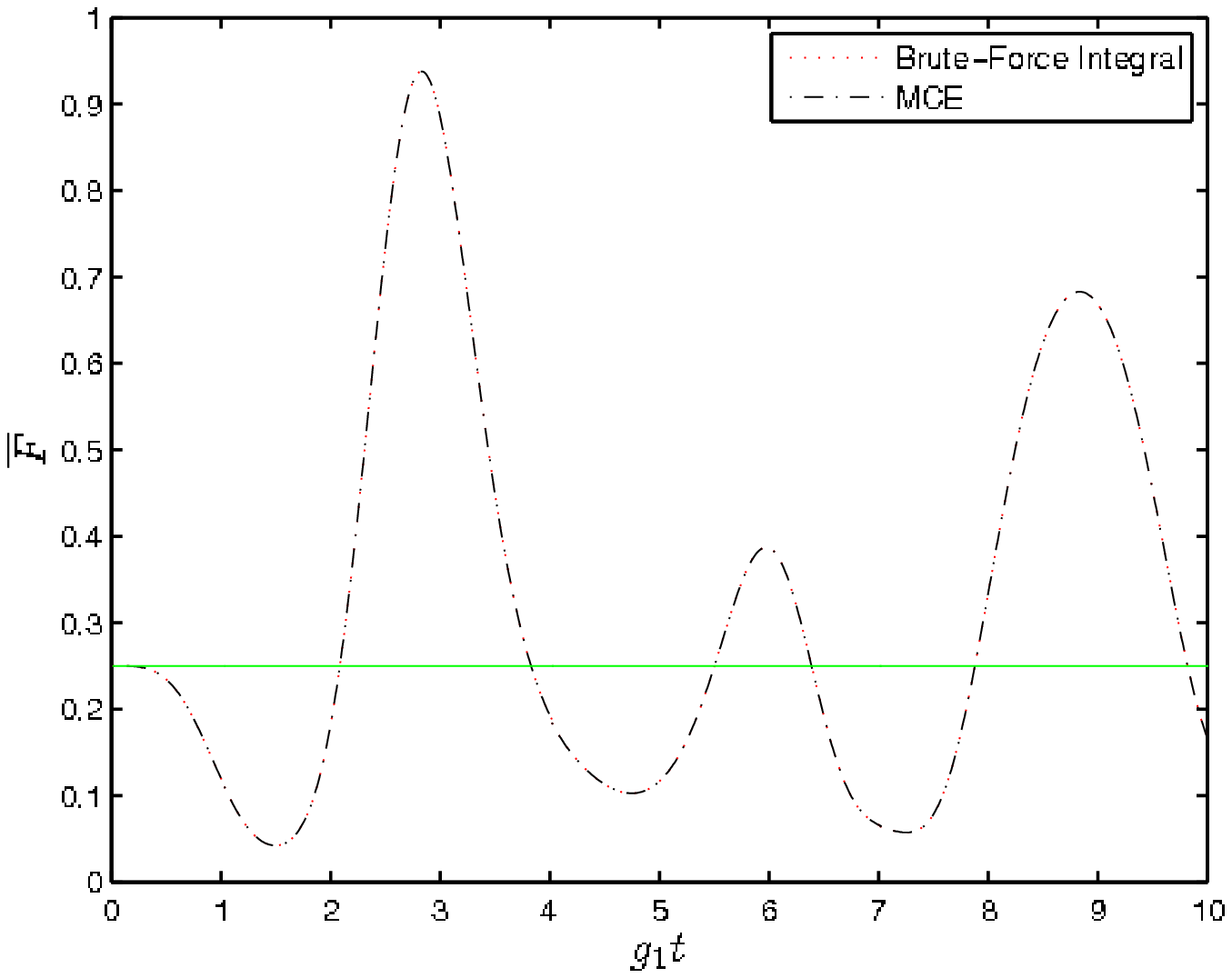}\label{1mode}}
\subfigure[]{\includegraphics[scale=0.26]{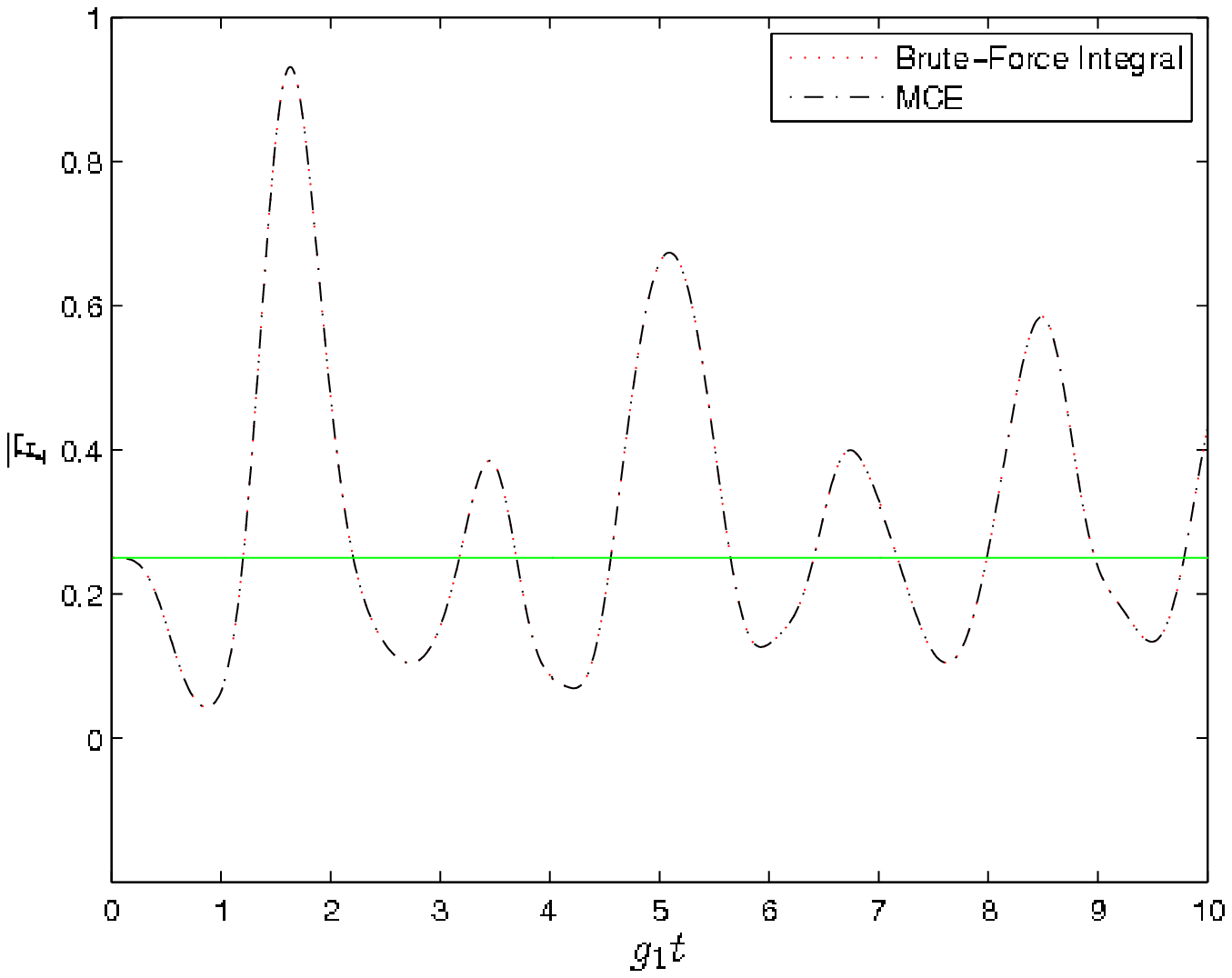}\label{3modes}}
\caption{Choi fidelity $F$ versus rescaled time, for $\hat{H}_{rw}$ with
$\varepsilon=\Delta=0$, $g_1=1$,
$g_2=1.9$, obtained at zero temperature
by MCE method (dot-dashed) and exact analytic integration (dotted)
for $M=1$ and $\omega_1=0.1$ (a), and $M=3$ and $\omega_m=0.1 m$ for $1\le m \le3$ (b).
The lines $F=0.25$ are reported for reference.}
\end{center}
\end{figure}

To define the gate fidelity $F$, we will make use of the
classic channel-state duality (Choi isomorphism) mapping linear quantum operations
over a Hilbert space ${\mathcal H}$ into quantum states on the Hilbert space ${\mathcal H}\otimes{\mathcal H}$ \cite{choi75}.
Turning to the two qubits Hilbert space ${\mathcal H}$ spanned by the basis states (\ref{alpha}-\ref{delta}), let us
define the maximally entangled fiducial state $\ket{\psi}$ (belonging to ${\mathcal H}^2$) as:
\be
\ket{\psi} = \frac12(\ket{{1}}\otimes \ket{{1}} + \ket{{2}}\otimes \ket{{2}}
+ \ket{{3}}\otimes \ket{{3}} + \ket{{4}}\otimes \ket{{4}}) \, .
\ee
For a  generic CP-map $\Omega$, the corresponding quantum state $\varrho_{\Omega}$ may be defined as
\be
\varrho_{\Omega} = (\Omega\otimes{\mathbbm 1})(\ket{\psi}\bra{\psi}) \; ,
\ee
where ${\mathbbm 1}$ is the identity map acting on ${\mathcal H}$.

Since the CZ gate is unitary, the quantum state $\varrho_{CZ}$ is bound to be pure:
$\varrho_{CZ} = \ket{\varphi_{CZ}}\bra{\varphi_{CZ}}$, with
\bea
\ket{\varphi_{CZ}} &=& \frac12\left(\ket{{1}}\otimes \ket{{1}} + \ket{{2}}\otimes \ket{{2}}
+ \ket{{3}}\otimes \ket{{3}} - \ket{{4}}\otimes \ket{{4}}\right) \nonumber \\
 &=& \frac{1}{2} \sum_{j=1}^{4} f(j)\left(\ket{{j}}\otimes\ket{{j}}\right) \; .
\eea
The state $\varrho_{\Gamma_{t,\beta}}$ corresponding to $\Gamma_{t,\beta}$ is instead given by
\be
\varrho_{\Gamma_{t,\beta}} = \frac{1}{4}\sum_{j,k=1}^{4} \Gamma_{t,\beta}\left(\ket{{j}}\bra{{k}}\right)\otimes\ket{{j}}\bra{{k}} \; .
\ee
We can then naturally define the CZ operation fidelity $F$ (which we shall informally refer to as `Choi fidelity') as the overlap
\be
F = \bra{\varphi_{CZ}}\varrho_{\Gamma_{t,\beta}}\ket{\varphi_{CZ}} =
\frac{1}{16}\sum_{j,k=1}^{4}  \bra{{j}}\Gamma_{t,\beta}\left(\ket{{j}}\bra{{k}}\right)\ket{{k}} \; .
\ee

The quantity $F$ captures, in one real number, a relevant facet of the quantum dynamics governing the two qubits.
Its relationship to quantum coherence is manifest in that if the off-diagonal elements between the basis
vectors of Eqs.~(\ref{alpha}-\ref{delta}) are set to zero, then one has $F\le1/4$. Any value of $F$ larger than $1/4$
is thus in a sense a signature of quantum coherence. More importantly, $F$ is also a measure of how well a coherent quantum
task can be performed and is also strictly related to the entanglement generated between the two qubits
(in that entanglement a perfect CZ gate would get entanglement equal to $1$ ebit for a properly chosen initial state).
Moreover, $F$, although partial to the chosen reference gate (CZ in this case) is completely independent of the initial state,
and represents a property of the {\em dynamics} alone.

Of course, we could have chosen more generic quantifiers like, for instance, the largest eigenvalue
of the operator $\varrho_{\Gamma_{t,\beta}}$, which would equal $1$ in the ideal case where the qubits undergo a unitary evolution and would quantify, in a sense, the overall coherence of the qubits' evolution. However, we deem such choices to be less informative with regard to the
applicative potential of a complex dynamics.

Given a potentially useful quantum dynamics,
the knowledge of $F$ is instead very desirable to possess.
Demonstrating the use of a numerical technique capable of providing one with reliable estimates of
$F$ in relevant situations is, in a nutshell, the aim of the current analysis.

\begin{figure}[t!]
\begin{center}
\includegraphics[scale=0.5]{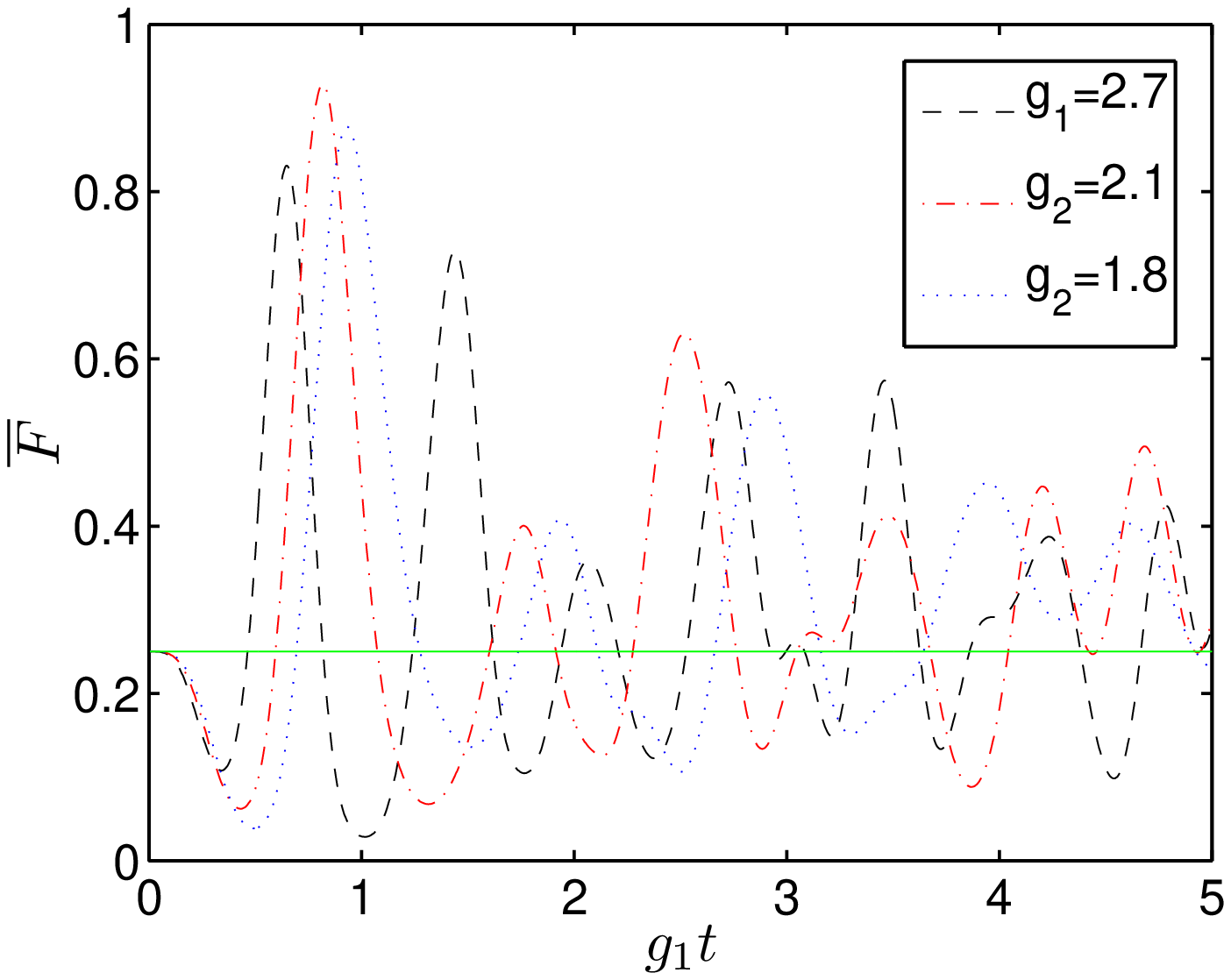}
\caption{MCE results for the Choi fidelity $F$ versus rescaled time, for $\hat{H}_{rw}$
with $\varepsilon=\Delta=0$, $g_1=1$,
$M=10$, $\omega_m=0.1 m$ for $1\le m\le 10$, zero temperature and different values of $g_2$.
The line $F=0.25$ is reported for reference.
\label{rwbetainf}}
\end{center}
\end{figure}
\begin{figure}[t!]
\begin{center}
\includegraphics[scale=0.5]{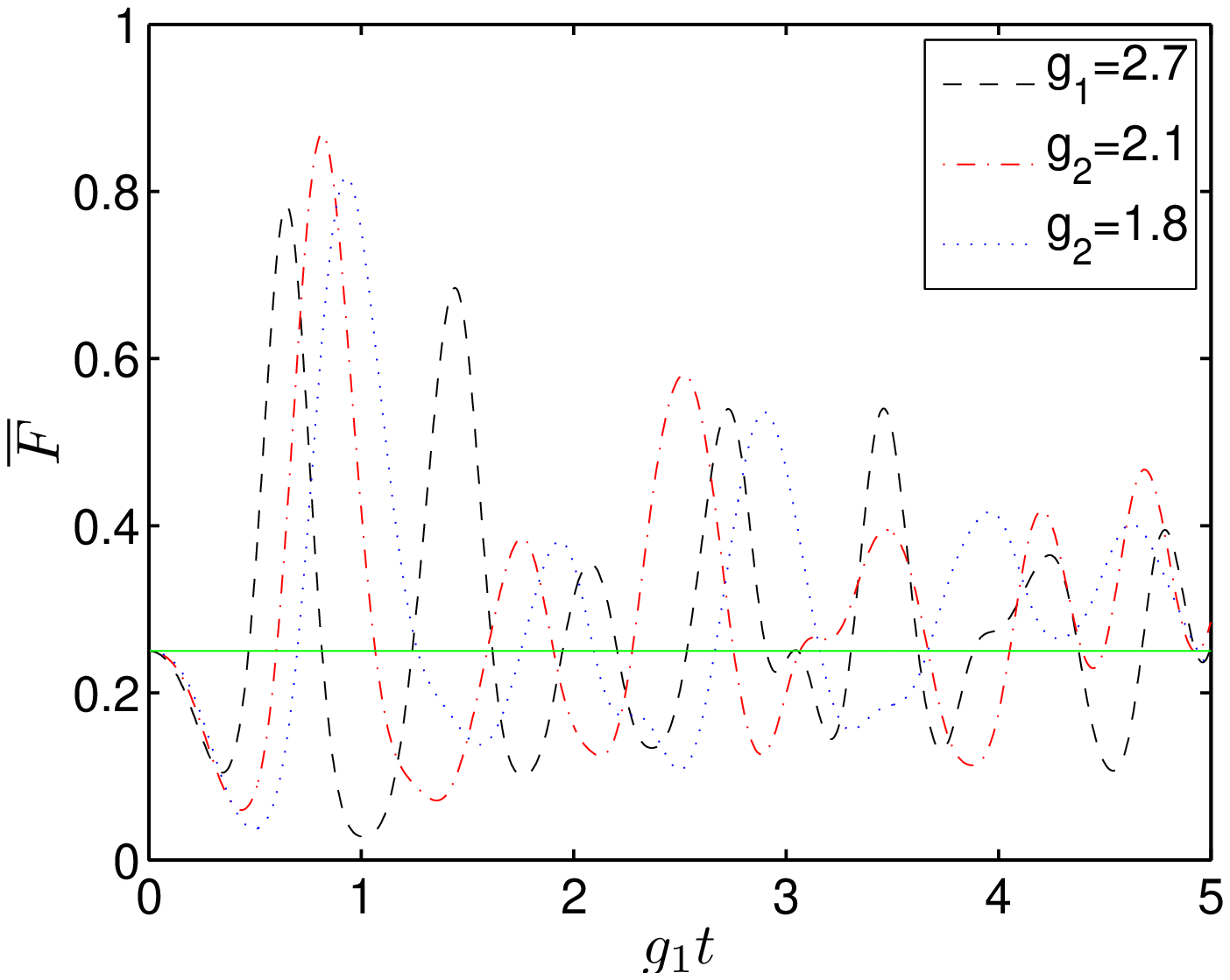}
\caption{MCE results for the Choi fidelity $F$ versus rescaled time, for $\hat{H}_{rw}$
with $\varepsilon=\Delta=0$, $g_1=1$,
$M=10$, $\omega_m=0.1 m$ for $1\le m\le 10$, $\beta=10$ and different values of $g_2$.
The line $F=0.25$ is reported for reference.
\label{rwbeta10_1}}
\end{center}
\end{figure}
\begin{figure}[t!]
\begin{center}
\includegraphics[scale=0.5]{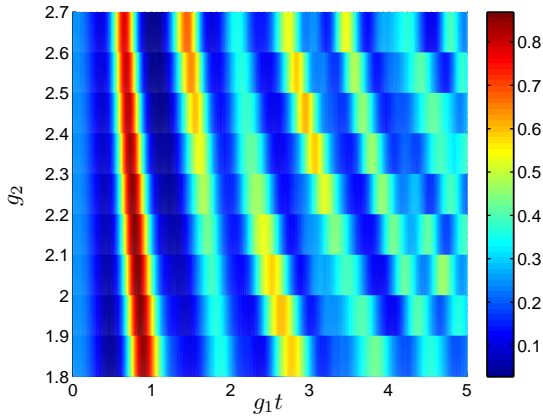}
\caption{MCE results for the Choi fidelity $F$ versus rescaled time, for $\hat{H}_{rw}$ with
$\varepsilon=\Delta=0$, $g_1=1$,
$M=10$, $\omega_m=0.1 m$ for $1\le m\le 10$, $\beta=10$ and different values of $g_2$ (red stands for higher values, blue for lower values).\label{rwbeta10_2}}
\end{center}
\end{figure}

\section{Choi fidelity of the CZ gate \label{results}}

Here, we will slightly deviate from the previously adopted notation by setting
$g^{(j)}_{l}=g_{j}$ for all $j$ and $l$.
It is important to remark that assuming equal couplings between each qubit and all the modes is in no
way essential
to our numerical approach.
Such an assumption can be -- and will be, in the following -- relaxed if needs be.

Let us then, to begin with, set the coupling between the first qubit and the field modes
$g_1$ to $1$, and essentially choose it as the unit of time. Let us also, until further notice, set $\varepsilon=\Delta=0$, $\beta\rightarrow\infty$ (zero temperature) and consider the rotating wave Hamiltonian $\hat{H}_{rw}$.
Note that the Hamiltonian $\hat{H}_{rw}$ with $\varepsilon=\Delta=0$ can be derived
from the full Hamiltonian $\hat{H}$ with $\Delta=0$ by switching to interaction picture and applying the rotating wave approximation: $\varepsilon$ can be thus be set to zero and each field frequency
$\omega_m$ is shifted as per $\omega_{m}\rightarrow \omega_{m}-2\varepsilon$.

\begin{figure}[t!]
\begin{center}
\includegraphics[scale=0.5]{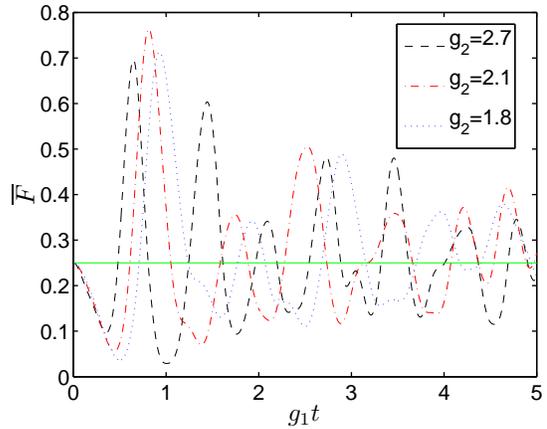}
\caption{MCE results for the Choi fidelity $F$ versus rescaled time, for $\hat{H}_{rw}$ with
$\varepsilon=\Delta=0$, $g_1=1$,
$M=10$, $\omega_m=0.1 m$ for $1\le m\le 10$, $\beta=5$ and different values of $g_2$.
The line $F=0.25$ is reported for reference.\label{rwbeta5_2}}
\end{center}
\end{figure}
\begin{figure}[t!]
\begin{center}
\includegraphics[scale=0.5]{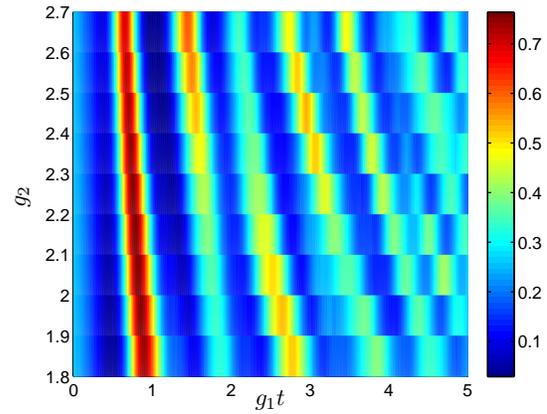}
\caption{MCE results for the Choi fidelity $F$ versus rescaled time, for $\hat{H}_{rw}$
with $\varepsilon=\Delta=0$, $g_1=1$,
$M=10$, $\omega_m=0.1 m$ for $1\le m\le 10$, $\beta=5$ and different values of $g_2$ (red stands for higher values, blue for lower values).\label{rwbeta5_2}}
\end{center}
\end{figure}

By exploiting the conservation of the number of excitations, the dynamics
governed by the Hamiltonian $\hat{H}_{rw}$ for $\Delta=0$ can be easily solved analytically.
The agreement between such analytical solutions and the MCE results has been tested for up to
ten modes and is excellent.
Figs.~\ref{1mode} and \ref{3modes} show such an agreement in terms of CZ Choi fidelity $F$ for
$g_2=1.9$ and, respectively, one mode with $\omega_1=0.1$ and three modes with $\omega_1=0.1$, $\omega_2=0.2$ and $\omega_3=0.3$.
An initial peak with fidelity larger than $0.9$ is immediately apparent: this peak will be the main
object of our investigation, for larger numbers of modes too.
For three modes, the peak appears at a time which is approximately reduced by a $\sqrt{3}$
factor with respect to the single mode case. This cooperative effect is confirmed for all number of modes up to $20$, and is simply due to the fact that the qubits are coupled to the field through the
mode $\frac{1}{\sqrt{M}}\sum_{m=1}^{M}a_{m}$, with an effective coupling which scales
like $\sqrt{M}$ (clearly, this is the consequence of assuming equal couplings with all modes).

Figs.~\ref{rwbetainf}-\ref{rwbeta5_2} depict a detailed analysis of the Choi fidelity for
$M=10$ bosonic modes with frequencies $\omega_m=0.1m$ for $1\le m\le10$, three different temperatures
($\beta\rightarrow\infty$, $\beta=10$ and $\beta=5$), and different values of the coupling $g_2$, scanned
over the range $1.8-2.7$. The zero temperature case (Fig.~\ref{rwbetainf}) shows how the dispersion
of quantum coherence among the field's degrees of freedom
affects the gate's fidelity (whose maximum is smaller than in the one- and three-modes cases),
although in the considered region of dynamical parameters the effect is not as pronounced as one could imagine.
The plots clearly show the detrimental effect of thermal fluctuations on coherent quantum effects, even at such relatively small temperatures.
In practice, this suggests severe constraints on the temperature for the observation of coherent quantum effects mediated by discrete vibrational modes (typically much more susceptible to thermal excitations than optical modes due to their lower frequency), considering that the highest temperature
accounted for is around $0.2 g_1$ in natural units.
Most importantly, we were able to determine the optimal value of the coupling $g_2$ with
respect to a vast range of values (much wider than what reported in the plots), in terms of
the maximal converged Choi fidelity $F$, and to establish
that $g_2\simeq 2.2$ yields the closest results to an ideal CZ gate.
Clearly, in practice, such couplings will not always be tunable at will, or possibly
only within a given windows of values: it is anyway remarkable to be able to identify optimal values
given a specific dynamical figure of merit.

\begin{figure}[t!]
\begin{center}
\includegraphics[scale=0.5]{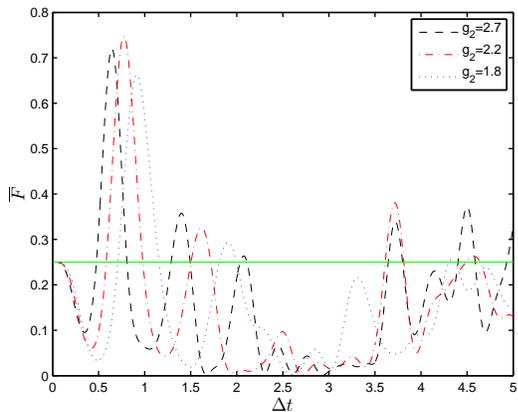}
\caption{MCE results for the Choi fidelity $F$ versus rescaled time
at zero temperature, for $\hat{H}_{rw}$ with
$\varepsilon=\Delta=1$, $g_1=1$,
$M=10$, $\omega_m=0.1 m$ for $1\le m\le 10$ and different values of $g_2$.
The line $F=0.25$ is reported for reference.\label{rwDelta1T0_1}}
\end{center}
\end{figure}
\begin{figure}[t!]
\begin{center}
\includegraphics[scale=0.5]{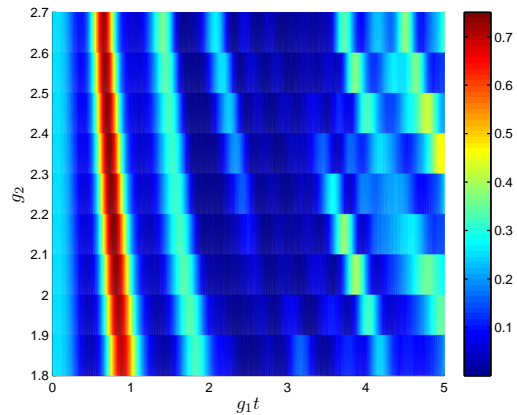}
\caption{MCE results for the Choi fidelity $F$ versus rescaled time at zero temperature,
for $\hat{H}_{rw}$ with
$\varepsilon=\Delta=1$, $g_1=1$,
$M=10$, $\omega_m=0.1 m$ for $1\le m\le 10$ and different values of $g_2$
(red stands for higher values, blue for lower values).\label{rwDelta1T0_2}}
\end{center}
\end{figure}
\begin{figure}[t!]
\begin{center}
\includegraphics[scale=0.5]{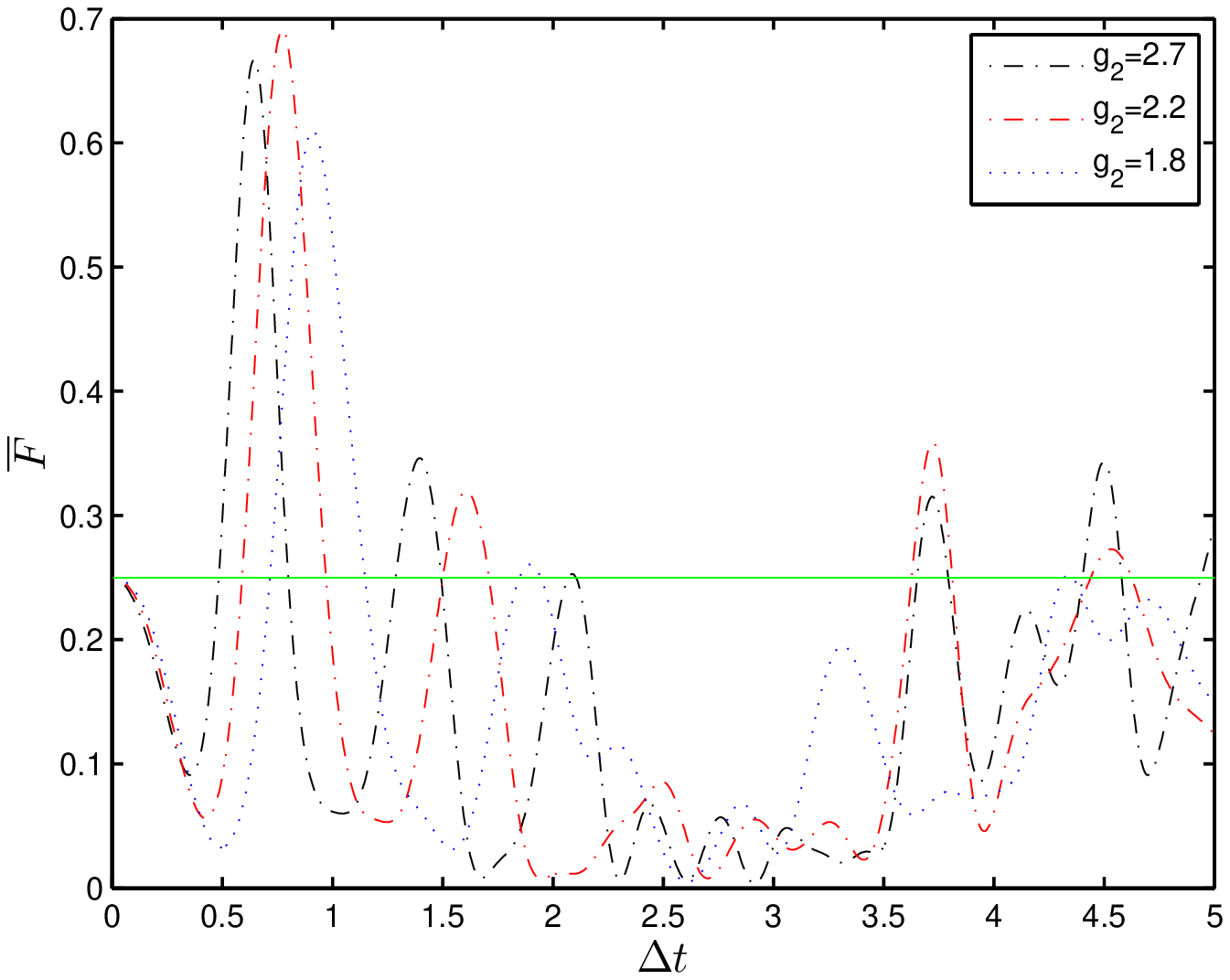}
\caption{MCE results for the Choi fidelity $F$ versus rescaled time, for $\hat{H}_{rw}$ with
$\varepsilon=\Delta=1$, $g_1=1$,
$M=10$, $\omega_m=0.1 m$ for $1\le m\le 10$, $\beta=10$ and different values of $g_2$.
The line $F=0.25$ is reported for reference.\label{rwDelta1beta10_1}}
\end{center}
\end{figure}
\begin{figure}[t!]
\begin{center}
\includegraphics[scale=0.5]{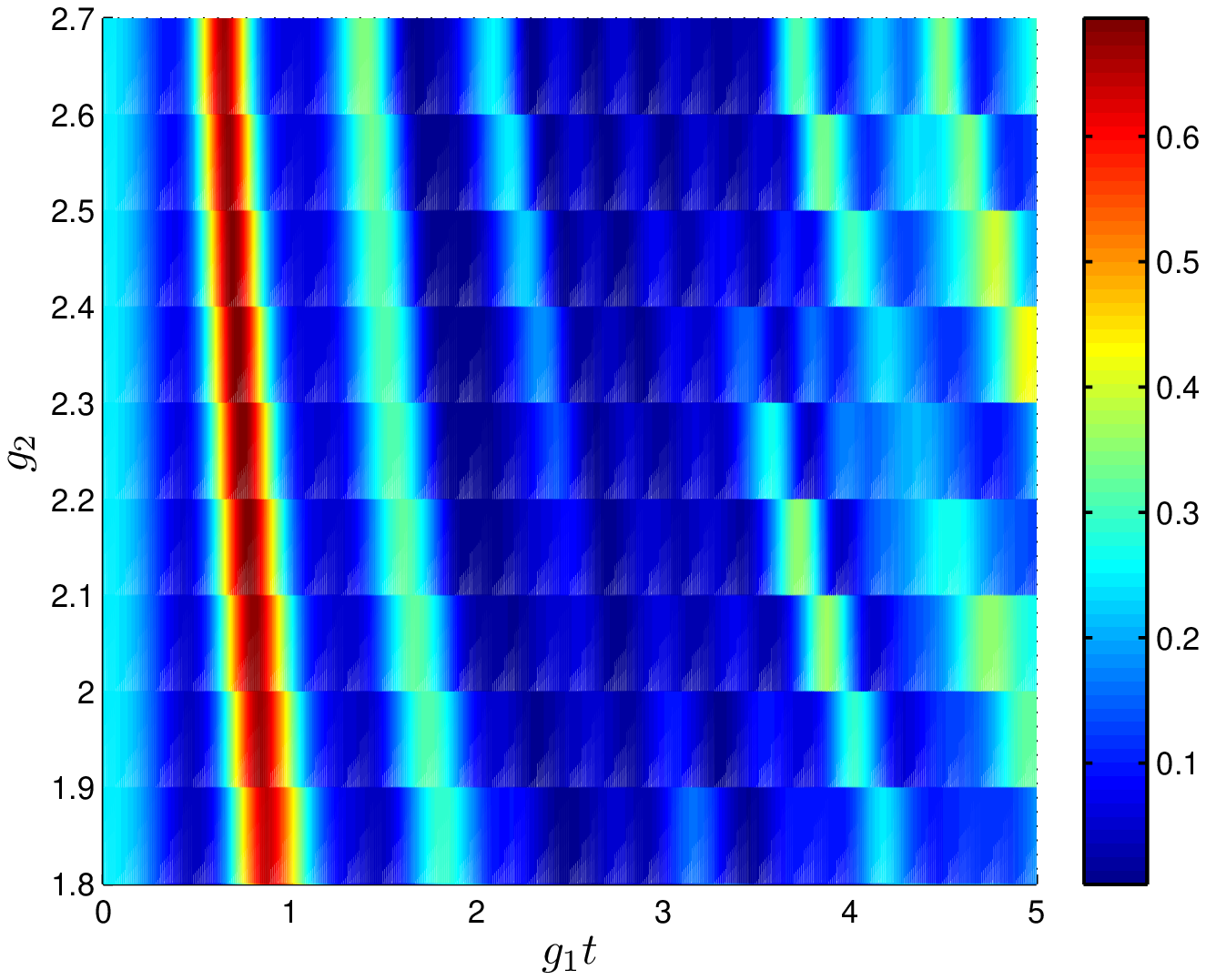}
\caption{MCE results for the Choi fidelity $F$ versus rescaled time, for $\hat{H}_{rw}$ with
$\varepsilon=\Delta=1$, $g_1=1$,
$M=10$, $\omega_m=0.1 m$ for $1\le m\le 10$, $\beta=10$ and different values of $g_2$
(red stands for higher values, blue for lower values).\label{rwDelta1beta10_2}}
\end{center}
\end{figure}

\begin{figure}[t!]
\begin{center}
\includegraphics[scale=0.5]{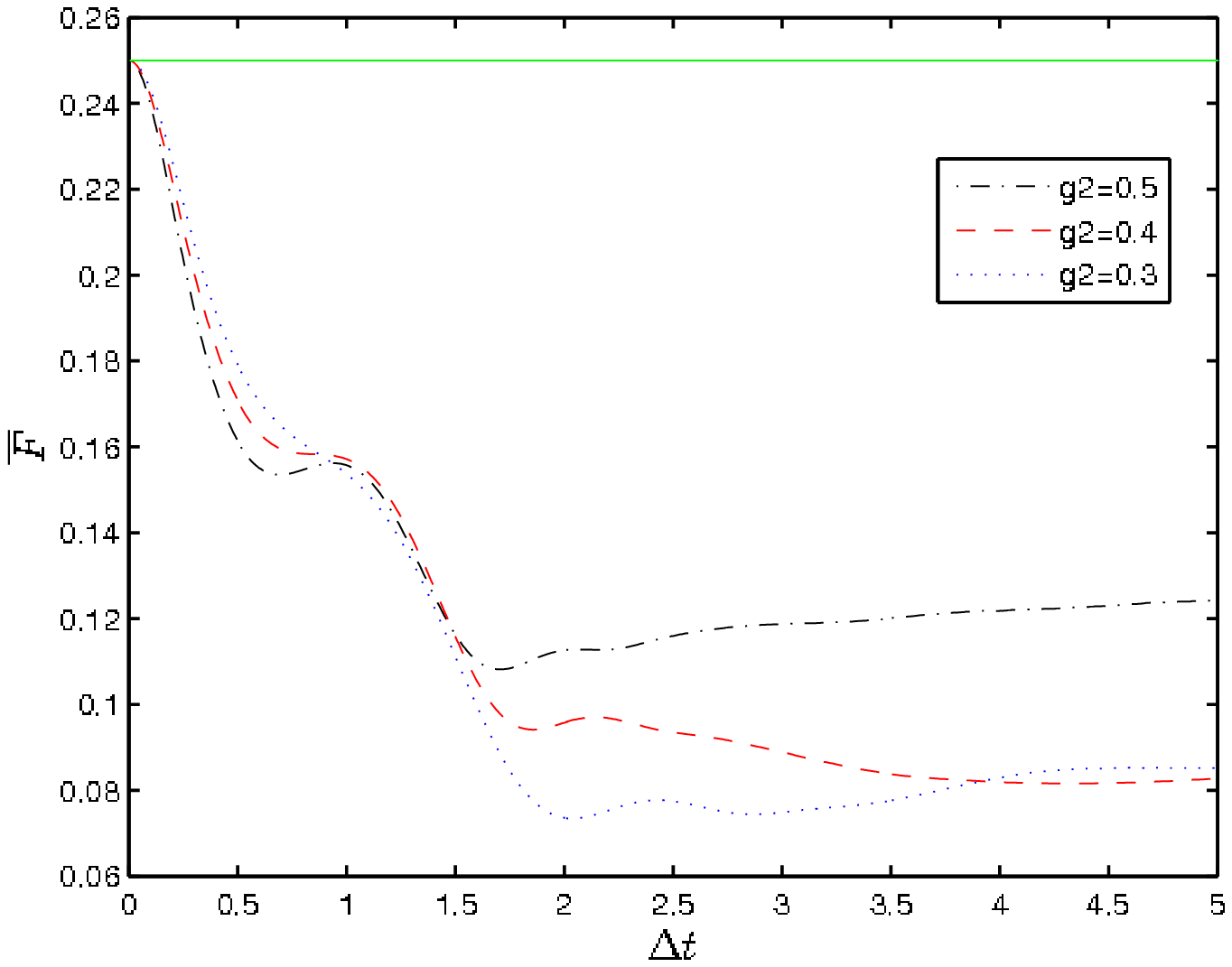}
\caption{MCE results for the Choi fidelity $F$ versus rescaled time
at zero temperature, for $\hat{H}$ with
$\varepsilon=\Delta=1$, $g_1=0.5$,
$M=10$, $\omega_m=0.1 m$ for $1\le m\le 10$ and different values of $g_2$.
The line $F=0.25$ is reported for reference.\label{sbT0}}
\end{center}
\end{figure}
\begin{figure}[t!]
\begin{center}
\includegraphics[scale=0.5]{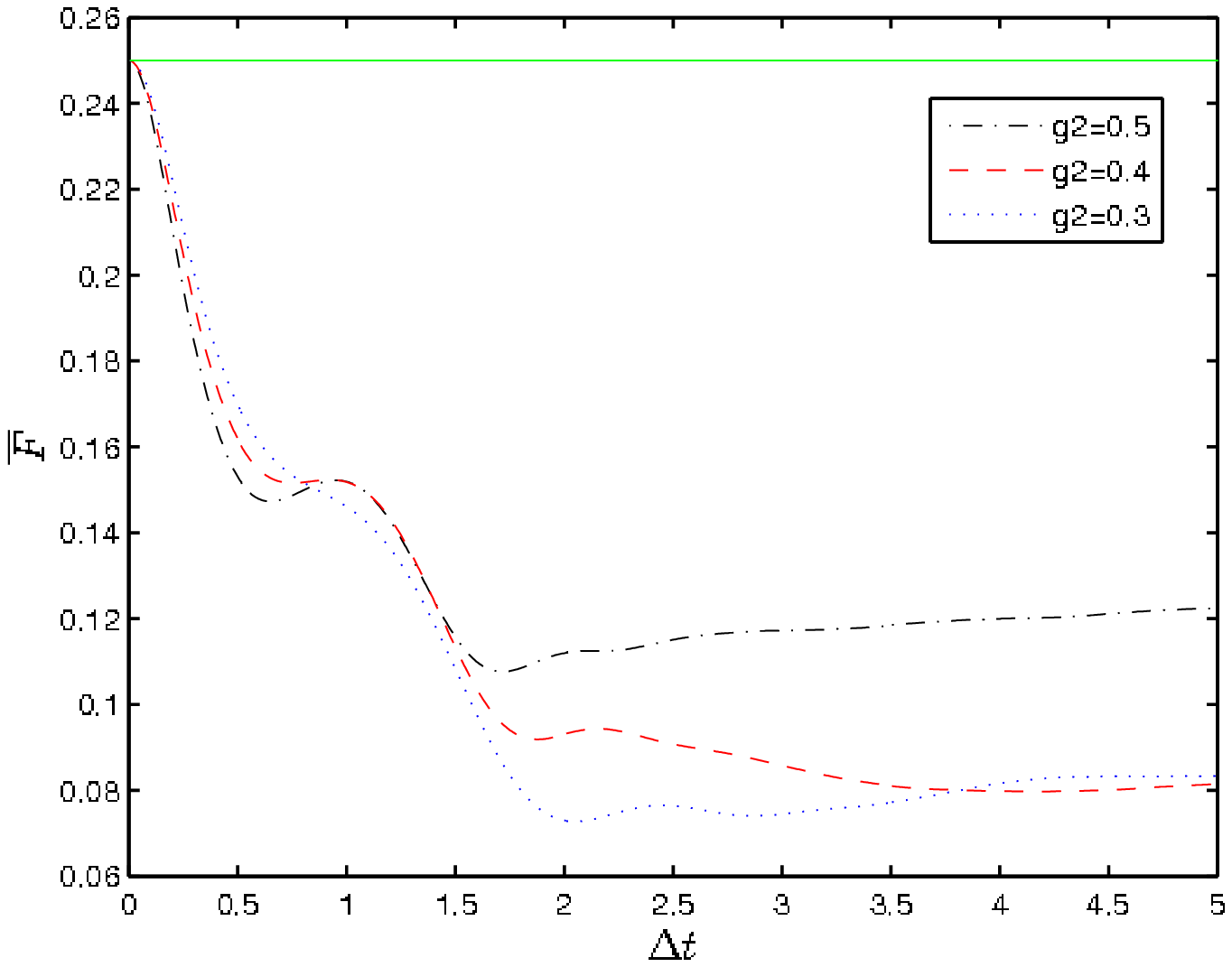}
\caption{MCE results for the Choi fidelity $F$ versus rescaled time, for $\hat{H}$ with
$\varepsilon=\Delta=1$, $g_1=0.5$,
$M=10$, $\omega_m=0.1 m$ for $1\le m\le 10$, $\beta=10$ and different values of $g_2$.
The line $F=0.25$ is reported for reference.\label{sbbeta10}}
\end{center}
\end{figure}

\begin{figure}[t!]
\begin{center}
\includegraphics[scale=0.5]{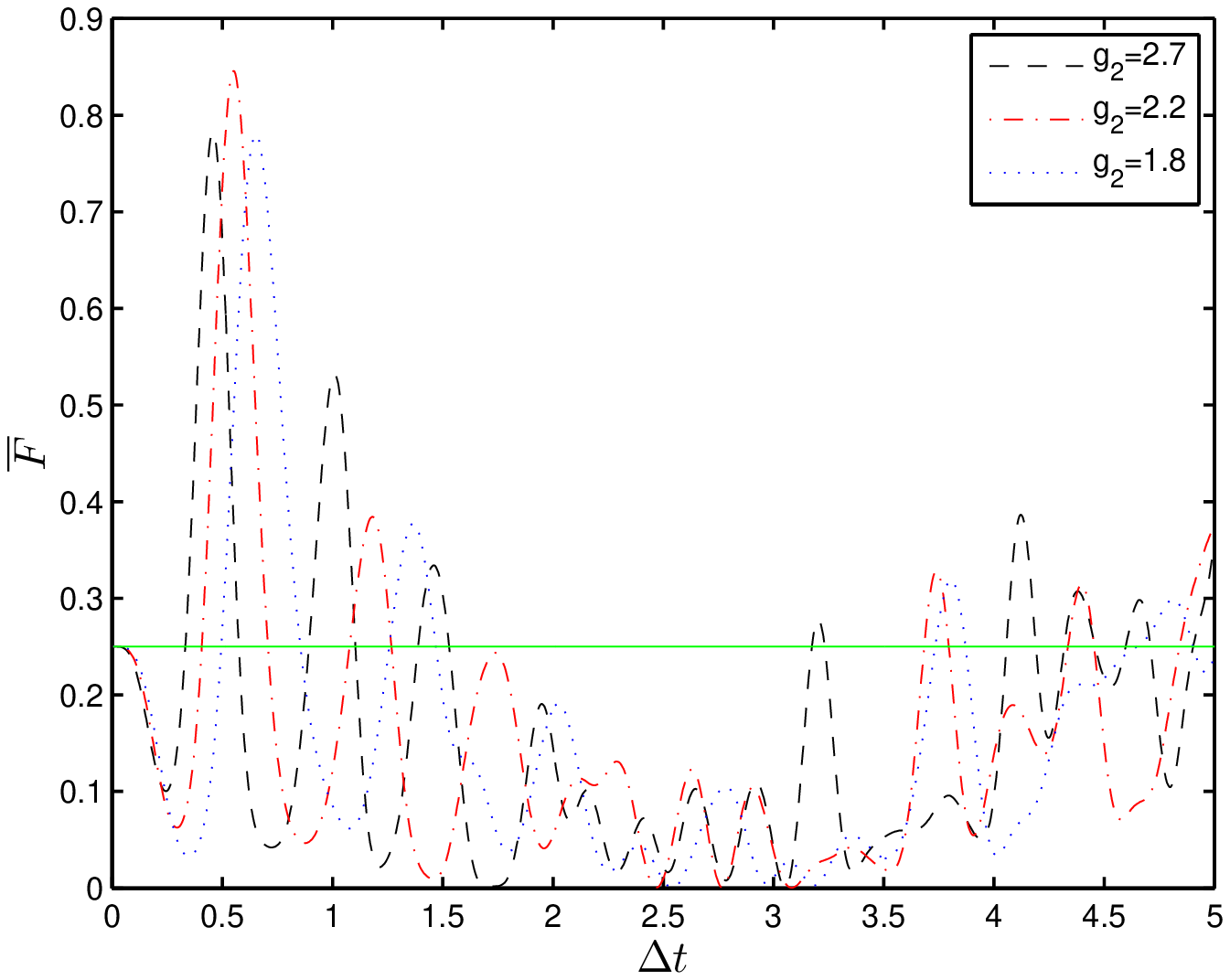}
\caption{MCE results for the Choi fidelity $F$ versus rescaled time
at zero temperature, for $\hat{H}_{rw}$ with
$\varepsilon=\Delta=1$, $g_1=1$,
$M=20$, $\omega_m=0.1 m$ for $1\le m\le 10$,
$\omega_m=0.1 (m-10)$ for $11\le m\le 20$ and different values of $g_2$.
The line $F=0.25$ is reported for reference.\label{doublerw}}
\end{center}
\end{figure}
\begin{figure}[t!]
\begin{center}
\includegraphics[scale=0.5]{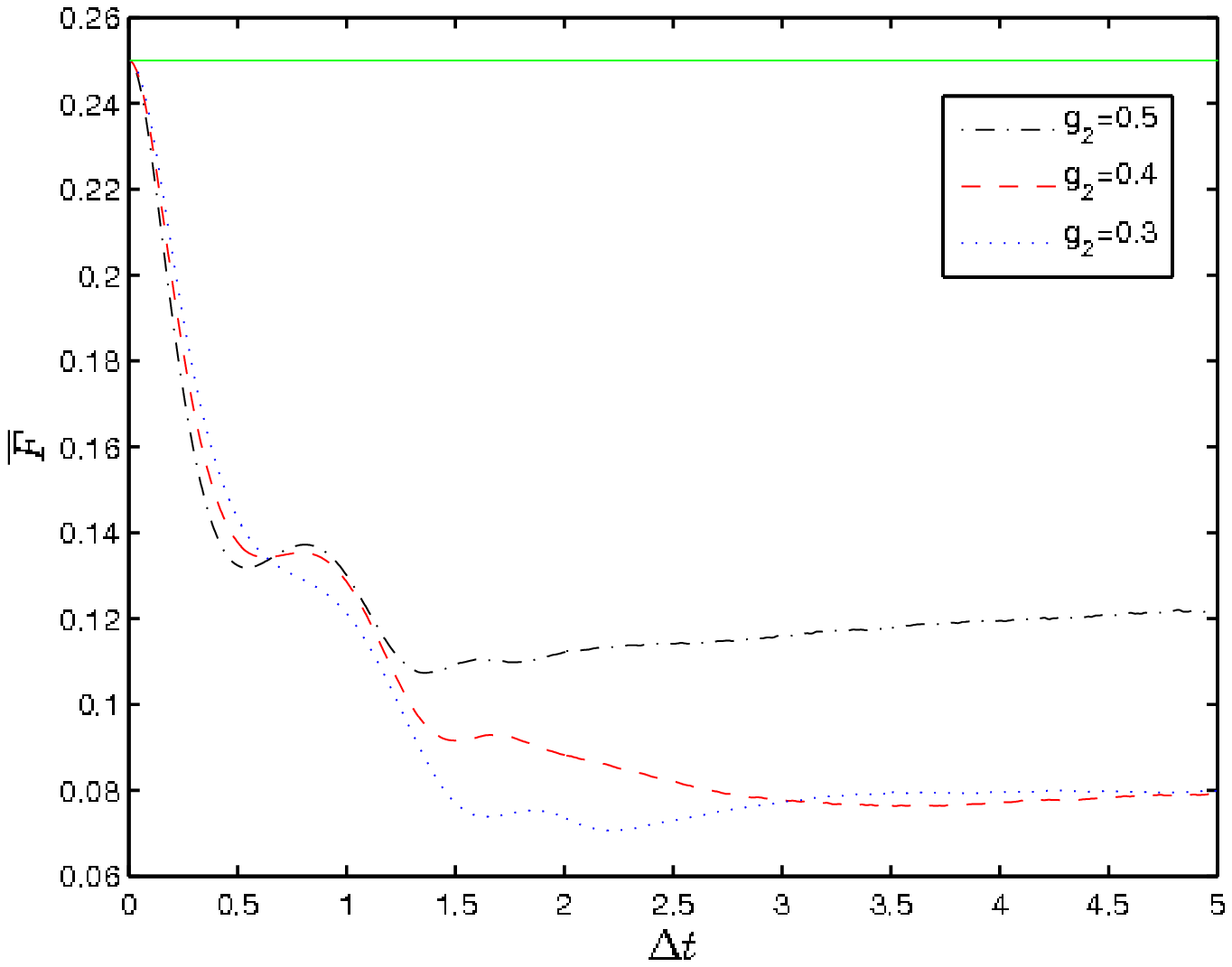}
\caption{MCE results for the Choi fidelity $F$ versus rescaled time
at zero temperature, for $\hat{H}$ with
$\varepsilon=\Delta=1$, $g_1=0.5$,
$M=20$, $\omega_m=0.1 m$ for $1\le m\le 10$,
$\omega_m=0.1 (m-10)$ for $11\le m\le 20$ and different values of $g_2$.
The line $F=0.25$ is reported for reference.\label{doublesb}}
\end{center}
\end{figure}

Let us now move on to a case which is not analytically treatable, with $\varepsilon=\Delta=1$
and $\omega_m=0.1m$ for $1\le m\le10$. As usual, we set $g_1=1$ and consider different values of
$g_2$ and different temperatures (zero temperature and $\beta=10$): the values of $F$
for such a configuration are displayed in Figs.~\ref{rwDelta1T0_1}-\ref{rwDelta1beta10_2}.
The initial peak in $F$ is still apparent, but the breaking of the phase invariance by the term $\Delta$
clearly degrades the quality of the gate, with a maximum Choi fidelity which is now around $0.7$, even
at zero temperature.

We now turn to the full spin-boson like Hamiltonian $\hat{H}$ (including the counter-rotating terms in the
qubits-field coupling), and consider the case $\varepsilon=\Delta=1$ and $\omega_m=0.1 m$
for $1\le m\le10$. The inclusion of the counter-rotating terms makes the simulation much
more challenging to run and converge. Roughly speaking, the main difficulty one encounters comes
down to the fact that the time-derivatives of the phase space positions of the basis grid,
determined by the Ehrenfest dynamics as per Eq.~(\ref{simpleton}), are much larger if the counter-rotating
terms are included. The time-dependent grid thus evolves much more rapidly in phase space and
is likely to leave the dynamically relevant region and accumulate substantial errors earlier.
We were however able to obtained well converged results by reducing the coupling $g_1$ to $0.5$,
which is still very far from the perturbative regime. Likewise, we scanned values of $g_2$ up to $0.5$.
Quite significantly -- as confirmed by Figs.~\ref{sbT0} and \ref{sbbeta10}, respectively
for zero temperature and $\beta=10$ -- we could not find any value of $g_2$ such that the Choi fidelity
of the CZ gate reaches 0.25. In fact, strong enough couplings are necessary to entangle the
two qubits on short enough time-scales but, with such strong couplings,
the counter-rotating terms heat the qubits up too quickly for coherent effects to take place,
at least in this region of parameters.
This heating overshadows the effect of thermal fluctuations in the field, which are barely noticeable
for $\beta=10$ (and are instead manifest in the rotating wave regime at the same temperature).

\begin{figure}[t!]
\begin{center}
\includegraphics[scale=0.5]{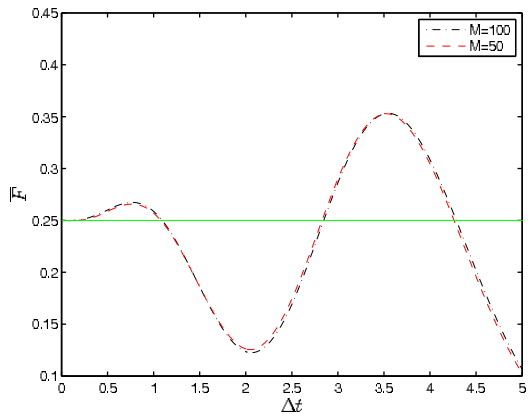}
\caption{MCE results for the Choi fidelity $F$ versus rescaled time
at zero temperature, for $\hat{H}$ with
$\varepsilon=\Delta=1$ and a common Ohmic bath with $\alpha=0.09$, $\omega_{c}=2.5$ and
different numbers of total bath modes.
The line $F=0.25$ is reported for reference.\label{ohmicsb}}
\end{center}
\end{figure}
\begin{figure}[t!]
\begin{center}
\includegraphics[scale=0.5]{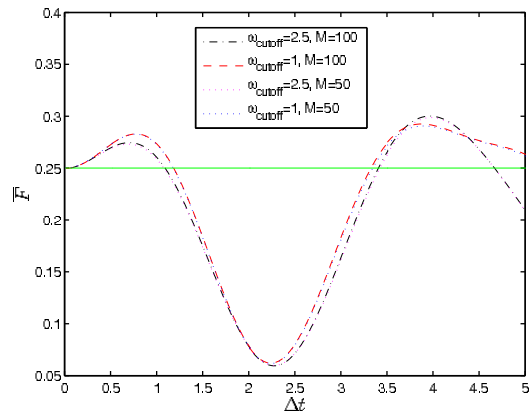}
\caption{MCE results for the Choi fidelity $F$ versus rescaled time
at zero temperature, for $\hat{H}_{rw}$ with
$\varepsilon=\Delta=1$ and a common Ohmic bath with $\alpha=0.09$, different cutoff frequencies
and different numbers of total bath modes.
The line $F=0.25$ is reported for reference.\label{ohmicrw}}
\end{center}
\end{figure}

\begin{figure*}[t!]
\begin{center}
\subfigure[]{\includegraphics[scale=0.34]{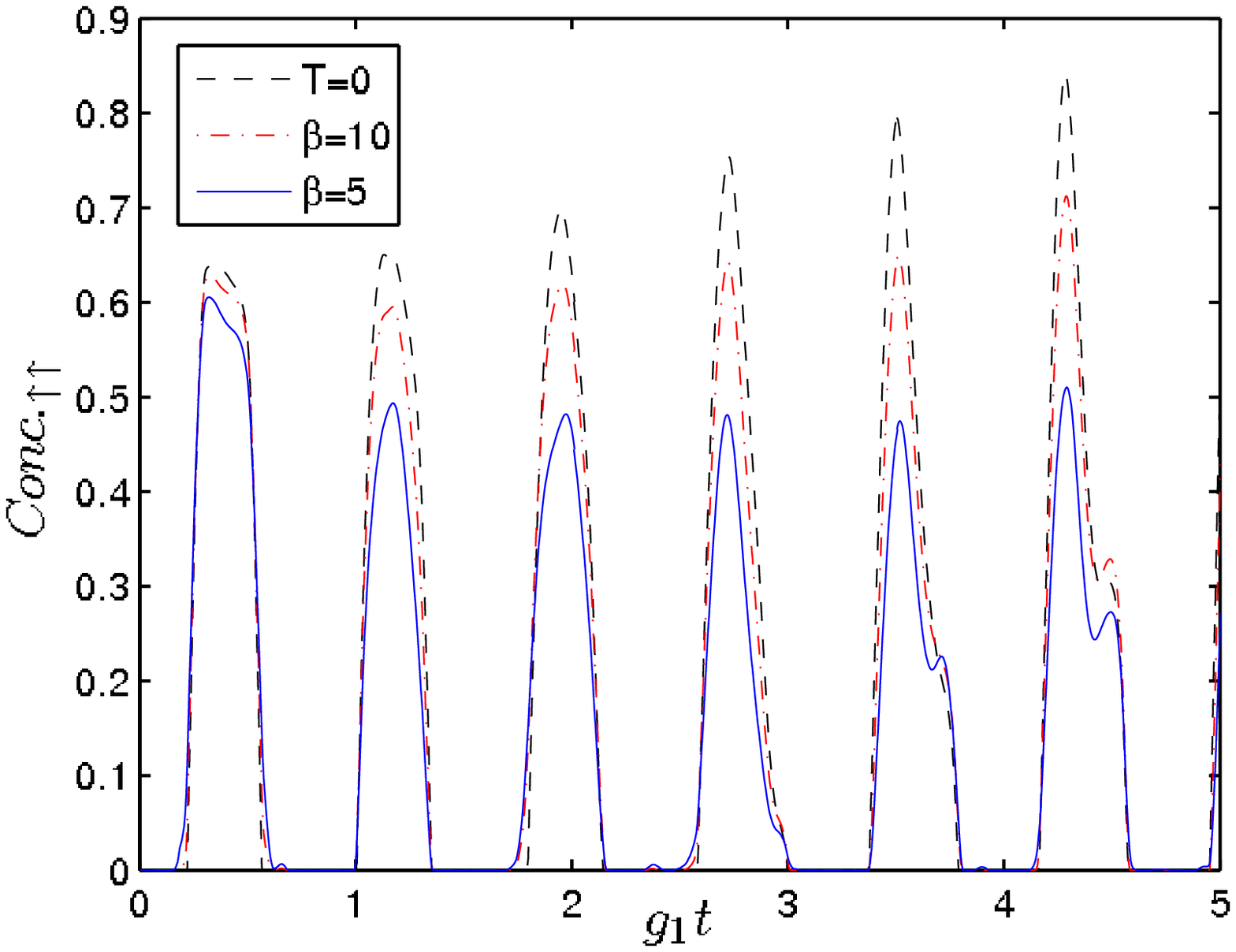}\label{conc1}}
\subfigure[]{\includegraphics[scale=0.34]{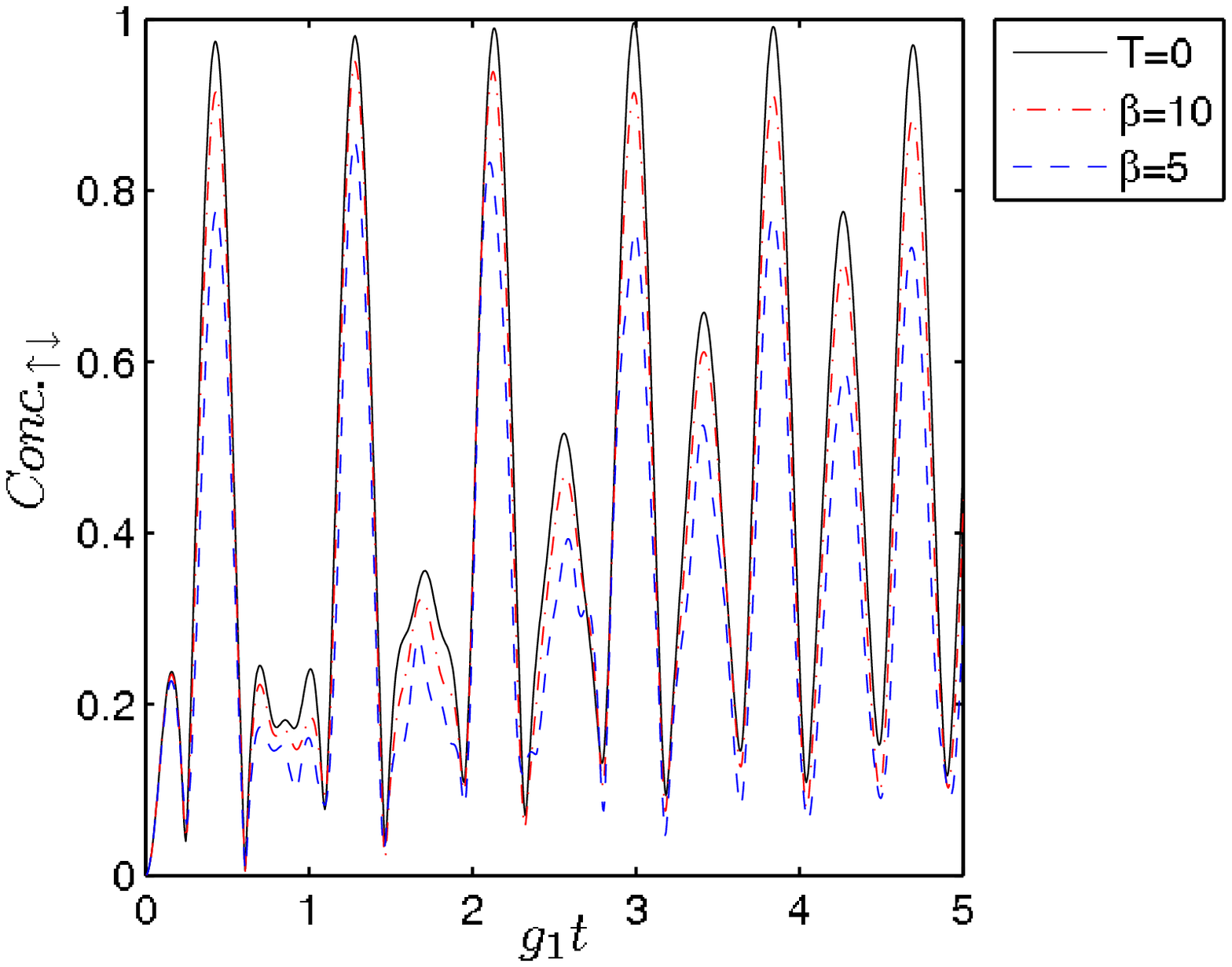}\label{conc2}}
\subfigure[]{\includegraphics[scale=0.34]{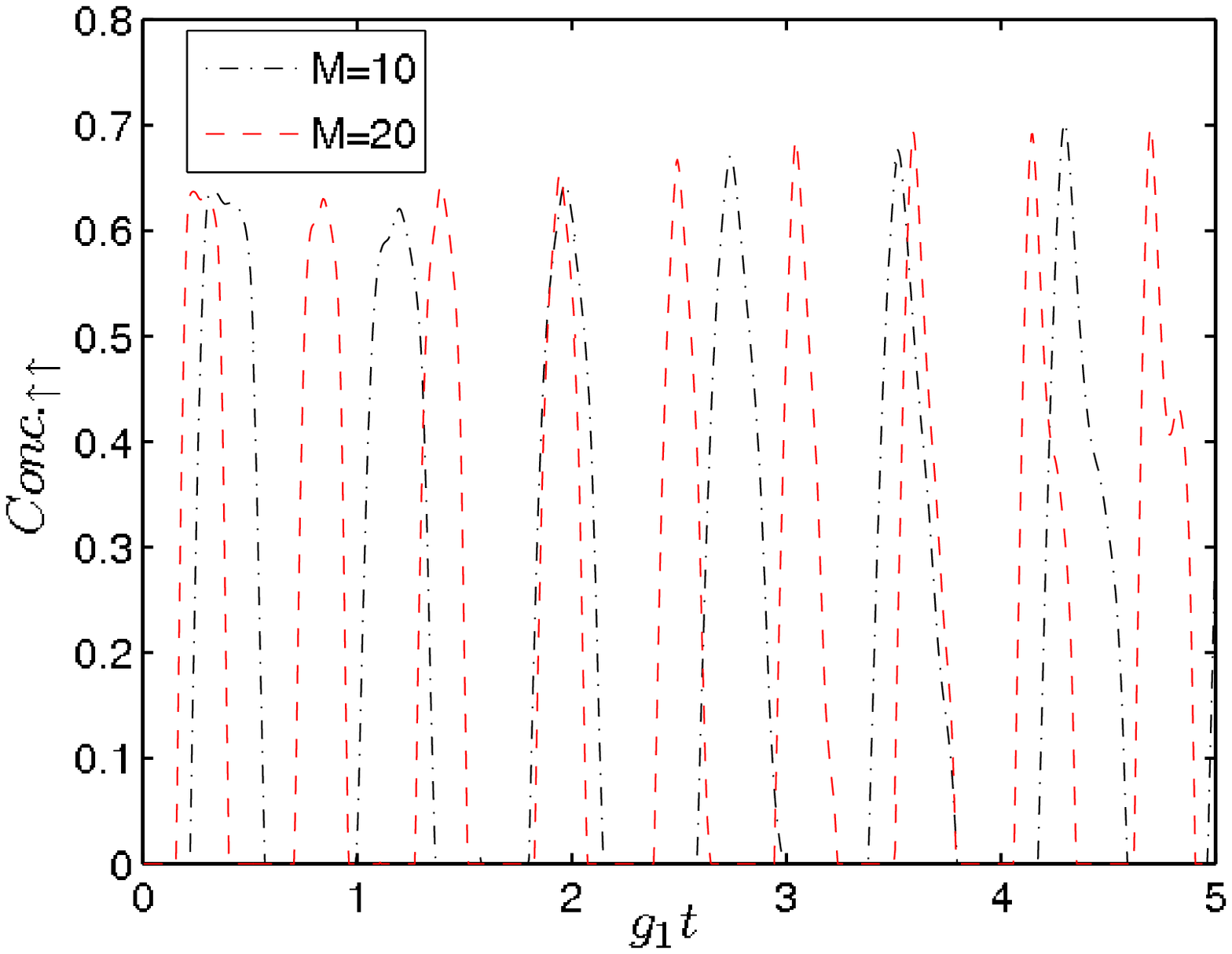}\label{conc3}}
\caption{MCE results for the concurrence versus rescaled time
at different temperatures for $\hat{H}_{rw}$ with $g_1=1$ and $g_2=2.1$.
In (a),  $\varepsilon=\Delta=0$, $M=10$ (with $\omega_m=0.1m$ for $1\le m\le10$)
and the initial state is $\ket{4}=\ket{\uparrow\uparrow}$; in (b), $\varepsilon=\Delta=0$, $M=10$
(with $\omega_m=0.1m$ for $1\le m\le10$)
and the initial state is $\ket{2}=\ket{\uparrow\downarrow}$;
in (c), $\varepsilon=\Delta=1$, the initial state is $\ket{2}=\ket{\uparrow\uparrow}$
and, respectively, $M=10$ (with $\omega_m=0.1m$ for $1\le m\le10$)
for the dash dotted line and $M=20$
(with $\omega_m=0.1m$ for $1\le m\le10$ and $\omega_m=0.1(m-10)$
for $11\le m\le20$) for the dashed line.}
\end{center}
\end{figure*}

In order to simulate the effect of a band of a 1-dimensional photonic band-gap medium,
where modes are usually doubly degenerate in frequency (since they can propagate in either spatial direction), we have also considered a case with $M=20$ modes, two for each equally spaced frequency.
All the other parameters have been kept as above, with $\varepsilon=\Delta=1$ and $g_1=1$
in the rotating wave case (Fig.~\ref{doublerw}), and $g_1=0.5$ in the full Hamiltonian
(Fig.~\ref{doublesb}). Comparing Fig.~\ref{doublerw} with Fig.~\ref{rwDelta1T0_1} shows that
the initial peak in Choi fidelity is still present: moreover, not only does it occur earlier by a factor $\sqrt{2}$
(as expected because of the cooperation between the modes due to the balanced coupling),
but it is also higher.
Contrary to common intuition, this example shows that
a larger number of mediating modes, in favourable dynamical configurations such as this,
can actually be advantageous for the implementation of locally coherent dynamics.
Note that,
for $M=20$ modes, we needed about $N=400$ coupled coherent states to achieve converged results.
This is as large a basis set as we used in this study.

\subsection{Zero temperature Ohmic spin-boson bath}

The notion of entangling separated systems and of distributing quantum coherence by interaction with common heat baths or
other incoherent means is well established in the quantum information and condensed matter communities,
and has been explored under a number of -- either more specific and applied
or more general and abstract -- viewpoints \cite{braun02,plenio02,kim02,cubitt03,benatti03,vorrath03,benatti05,oh06,solenov06,solenov07,an07,choi07,contreras08,dara,benatti09}.
However, the problem of studying the non-perturbative interaction of two qubits with a common bath is still in general a difficult one.
Thus, to provide the reader with further evidence of the versatility and power of our approach, we also report the application of the MCE method to the controlled-Z Choi fidelity for the case of the spin boson
Hamiltonian $\hat{H}$, with $\varepsilon=\Delta=1$, and both qubits interacting with a common bath
at zero temperature and
with Ohmic spectral density $J(\omega)$ given by
\be
J(\omega) = \frac2\pi \alpha \omega {\rm e}^{-\frac\omega\omega_c} \; ,
\ee
where $\alpha$ is the Kondo parameter, which we fix at $0.09$, and $\omega_c$ is a cutoff frequency.

We use a standard approach to discretise the bath, which has already been proven very reliable for single spin Ohmic spin-boson systems \cite{shalashilin09}. In particular, the frequencies and coupling strengths are chosen as follows:
\be
\omega_m = -\omega_c \ln\left[1-\frac{m\left(1-{\rm e}^{-\frac{\omega_{max}}{\omega_c}}\right)}{M}\right] \, ,
\ee
\be
g^{1}_m = g^{2}_m = \sqrt{\frac{\omega_m\alpha\omega_c\left(1-{\rm e}^{-\frac{\omega_{max}}{\omega_c}}\right)}{2M}} \; ,
\ee
where $\omega_{max}$ is a free parameter of the numerics, which we will
converge our results against (in that we choose it large enough to obtain converged results).
In particular, we choose $\omega_{max}=12.5$ for $\omega_c=2.5$ and $\omega_{max}=6$
for $\omega_{c}=1$. Also, the coupling strengths $g^{j}_m$ are defined as in Eqs.~(\ref{Hsb})
and (\ref{Hrw}).

Fig.~\ref{ohmicsb} shows the convergence of our results
for $\omega_c=2.5$ and the full Hamiltonian $\hat{H}$
in terms of the total number of modes in the bath ($M=50$ and $M=100$).
Quite interestingly, if the counter-rotating terms are included we obtain larger gate fidelities
when mimicking a bath than for a smaller set of discrete bus frequencies.
Off resonant modes are more influential in the full Hamiltonian and seems to be captured faithfully
by our method.

In Fig.~\ref{ohmicrw}, instead, we report results for the rotating wave Hamiltonian $\hat{H}_{rw}$
and different cutoff frequencies and number of bath modes. In this case, the effect of the
counter-rotating terms is rather limited. Also, the influence of larger cutoff frequencies at longer times is
clearly visible in the plot.

\section{Entanglement generation \label{ent}}

Typically, a large Choi fidelity for the (entangling) CZ gate corresponds to the
generation of substantial entanglement between the two qubits.
To support this statement, we report here a brief study on the entanglement generated between
two qubits. As an entanglement quantifier, we adopt the concurrence, an entanglement monotone
that can be easily calculated for a system of two qubits \cite{wootters98}.
Figs.~\ref{conc1}, \ref{conc2} and \ref{conc3} show the concurrence versus rescaled time for
the rotating wave Hamiltonian $\hat{H}_{rw}$ with
different initial states, temperatures, dynamical parameters and number of modes.
The degradation of quantum entanglement due to temperature is apparent
(Figs.\ref{conc1} and \ref{conc2}), along with the
speed up in the entanglement generation induced by a doubling of the modes (Fig.~\ref{conc3}).

It is also worth noticing that we did not find any region of parameters where entanglement between the two qubits
is generated for the full Hamiltonian $\hat{H}$ and $M=10$, thus mirroring our failure in obtaining
Cz fidelities larger than $0.25$.


\section{Conclusions \label{conc}}

We have presented an extensive numerical study, based
on Multi-Configurational Ehrenfest trajectories, on the dynamics of
two qubits interacting with a common set of bosonic field modes,
obtaining converged results for the Choi fidelity of an entangling
CZ gate between the qubits for a rather wide range of Hamiltonian parameters
and field temperatures, which cannot be covered by perturbation theory
or other approximate approaches.
We thus demonstrated the capability of tracking, analysing in detail,
and even optimising with respect of certain ranges of some parameters,
specific aspects of the coherent quantum dynamics of the qubits.

We were able to properly take into account the effect of finite bath's temperatures on the
reduced dynamics of the qubits, and also to highlight some counterintuitive features
related to the scaling of coherent signatures with the number of
field modes (which we varied over the range $1-100$), showing that at times more mediating modes can actually be advantageous for the distribution of quantum coherence.

The main limitations of our approach lie in the difficulty of handling counter-rotating
qubit-field coupling terms in the strong coupling regime ({\em i.e.}, when the coupling strengths are
comparable to the inherent dynamical frequencies of the qubits).
Even in such instances, we could however reach convergence by somewhat limiting the range of the
coupling strengths.

Within such limitations, the MCE approach has hence been established as a powerful
tool for the detailed study of complex quantum dynamics even with relatively limited resources
(desktop computers), typically for systems where discrete sets of up to 100 field modes
are involved.

\subsection*{Acknowledgments}

We thank Hannu Wichterich, Rui Zhang and LianHeng Tong for their help during the writing of the code.
SYY has been supported by a KC Wong Scholarship while pursuing this research.
DS acknowledges financial support from EPSRC, through grants
EPSRC EP/I014500/1 and NSF/EPSRC EP/J001481/1.
AS thanks the Central Research Fund of the University of London for financial support.


\begin{figure}[t!]
\begin{center}
\subfigure[Norm versus rescaled time]{\includegraphics[scale=0.26]{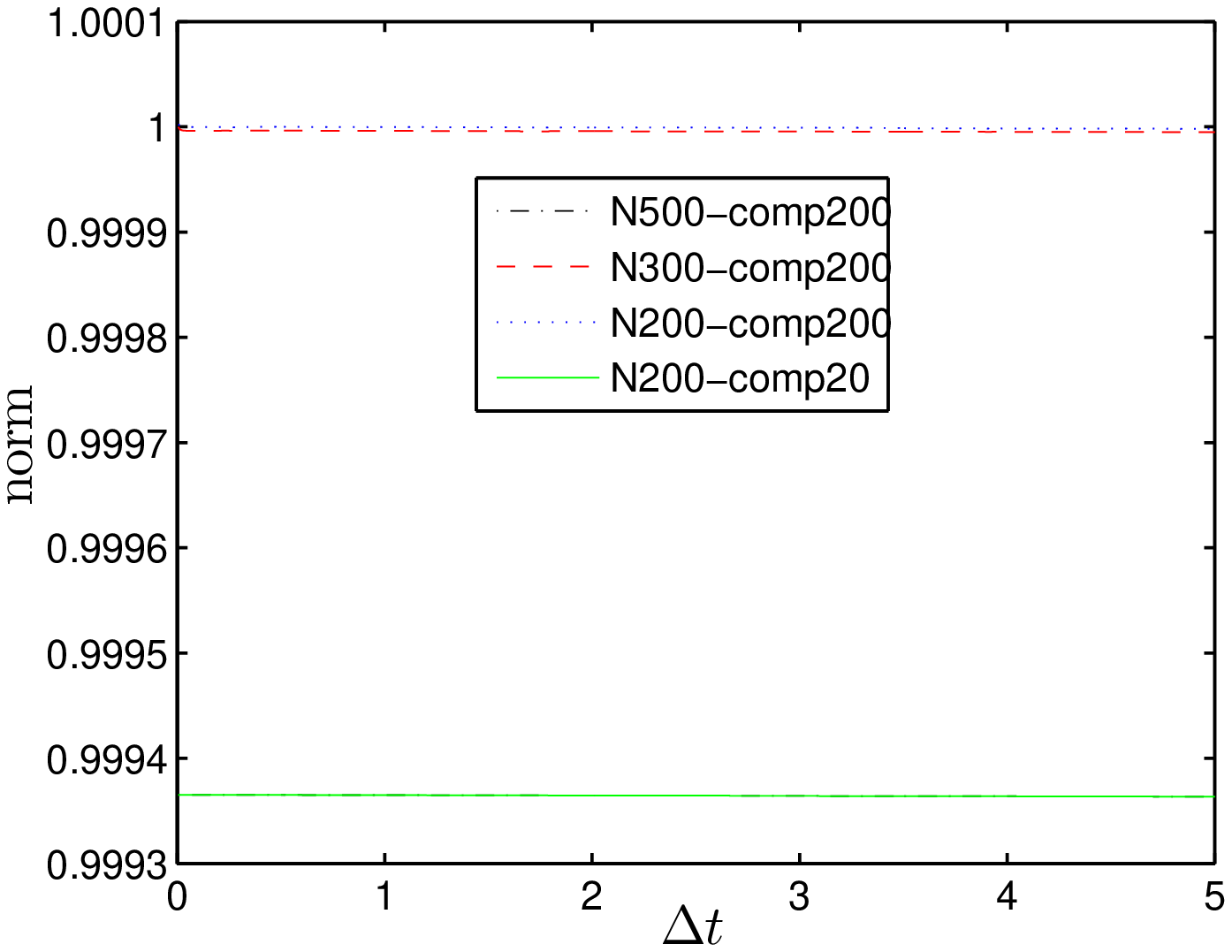}
\label{normrw}}
\subfigure[Energy versus rescaled time]{\includegraphics[scale=0.26]{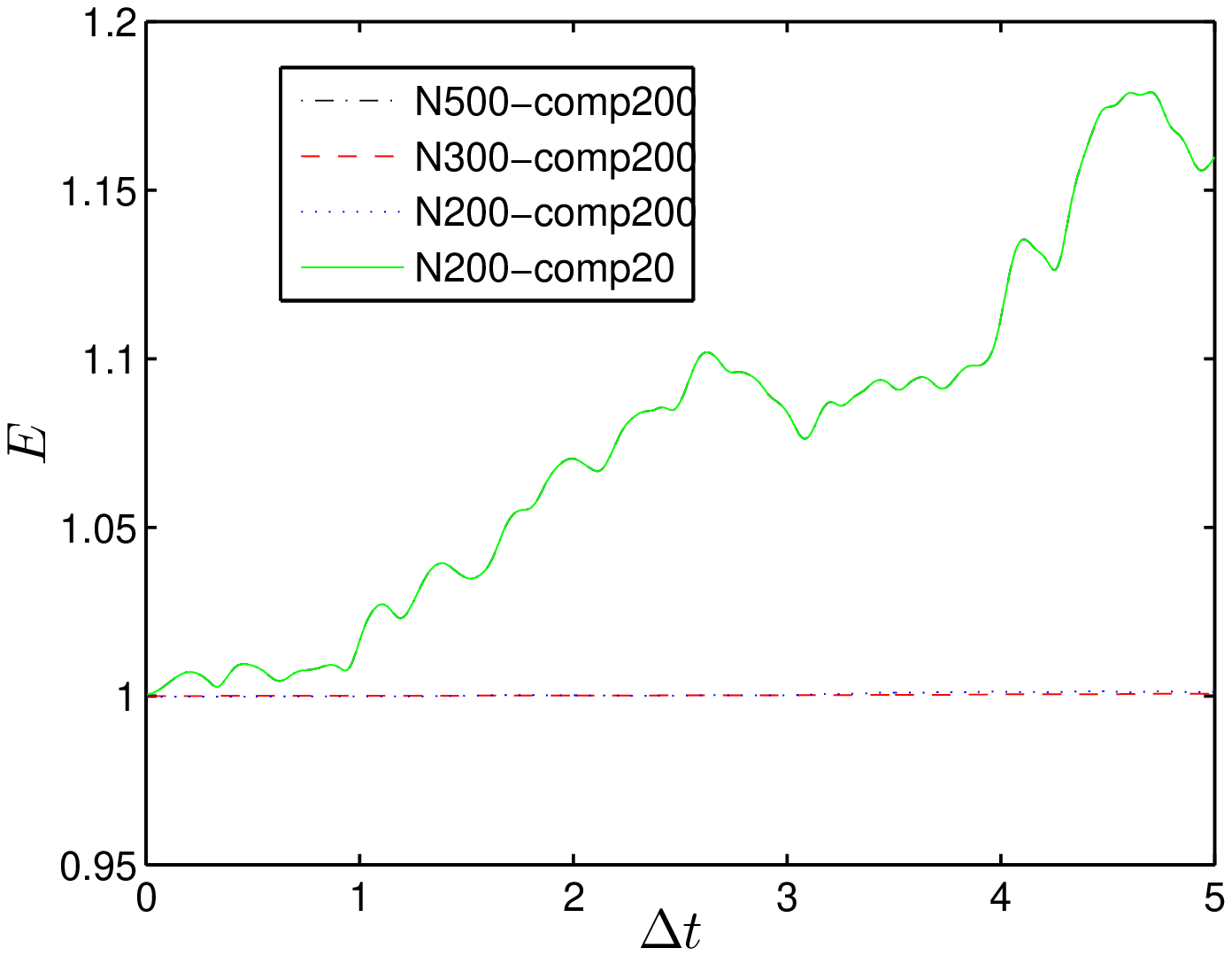}
\label{enerw}}
\caption{Norm and expectation value of the energy for MCE results at zero temperature,
for $\hat{H}_{rw}$ with
$\varepsilon=\Delta=g_1=1$, $g_2=2.7$, $M=10$ (with $\omega_m=0.1m$ for $1\le m\le10$) and different values of
$N$ and $comp$.}
\end{center}
\end{figure}

\begin{figure}[t!]
\begin{center}
\subfigure[$\varrho_{11}$ versus rescaled time]{\includegraphics[scale=0.26]{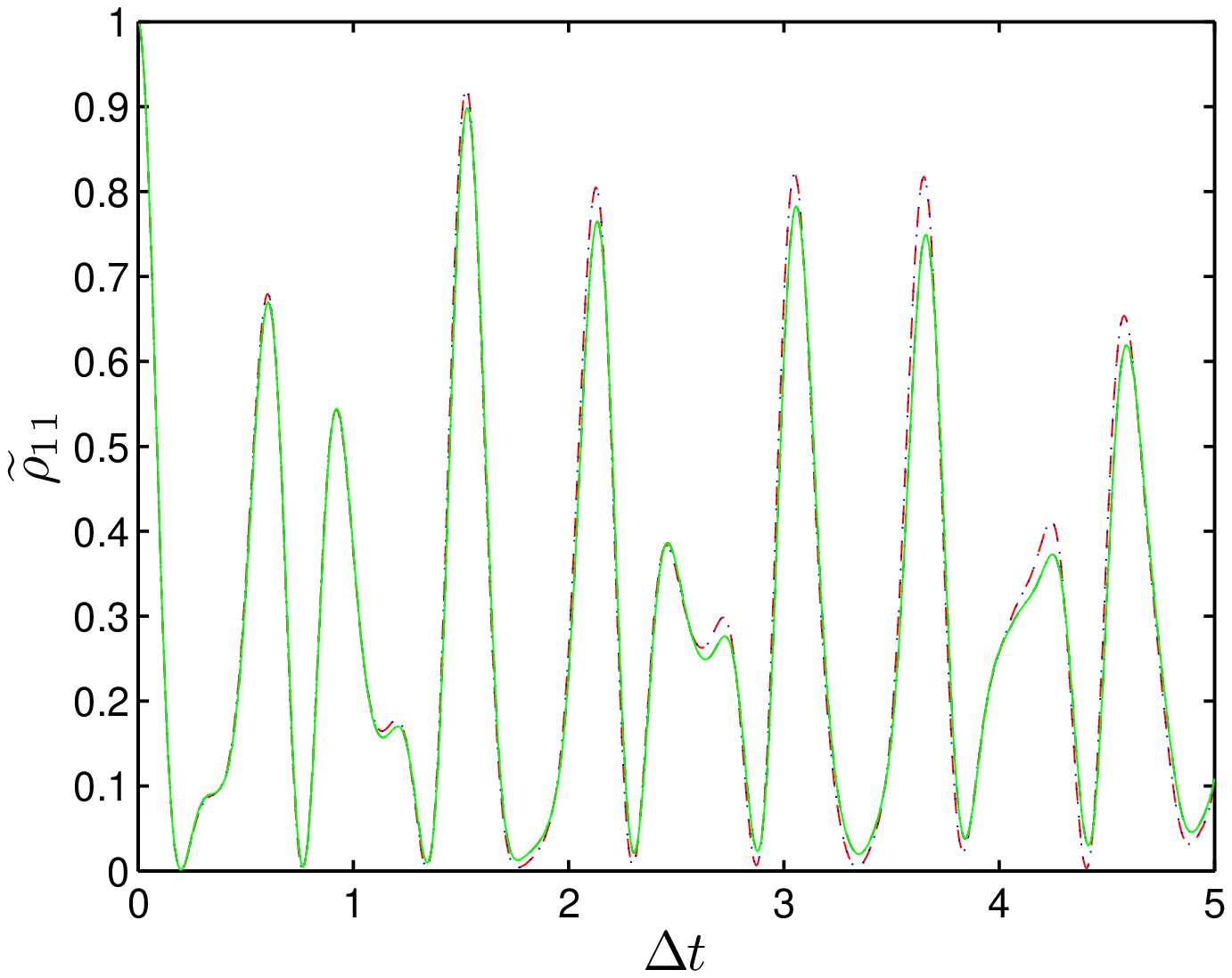}
\label{rho11rw}}
\subfigure[$\varrho_{13}$ versus rescaled time]{\includegraphics[scale=0.26]{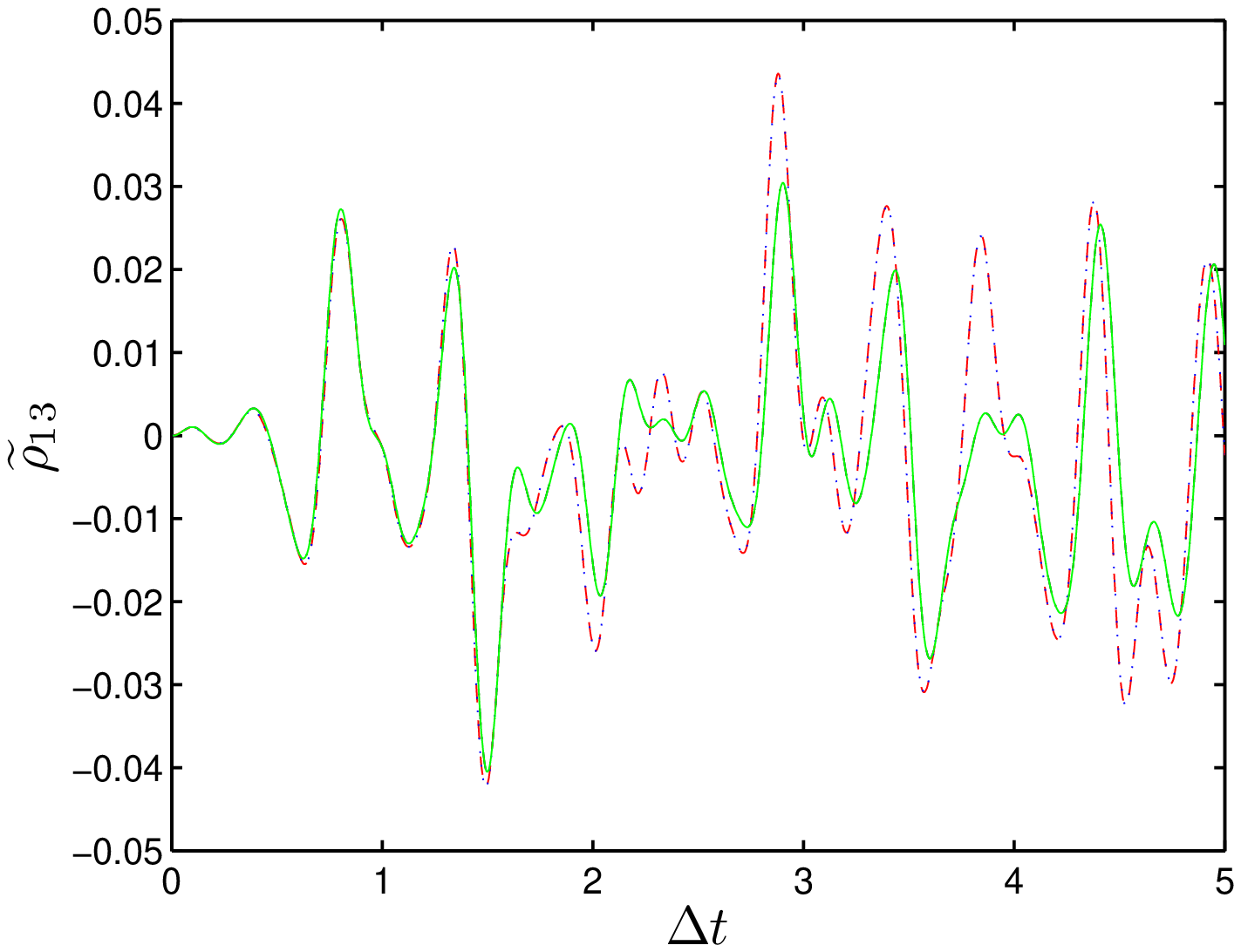}
\label{rho13rw}}
\caption{Entries of the qubits' density matrix $\varrho_{11}$ and $\varrho_{13}$ for MCE results at zero temperature,
for $\hat{H}_{rw}$ with
$\varepsilon=\Delta=g_1=1$, $g_2=2.7$, $M=10$ (with $\omega_m=0.1m$ for $1\le m\le10$) and different values of
$N$ and $comp$.}
\end{center}
\end{figure}
\begin{figure}[t!]
\begin{center}
\subfigure[Norm versus rescaled time]{\includegraphics[scale=0.26]{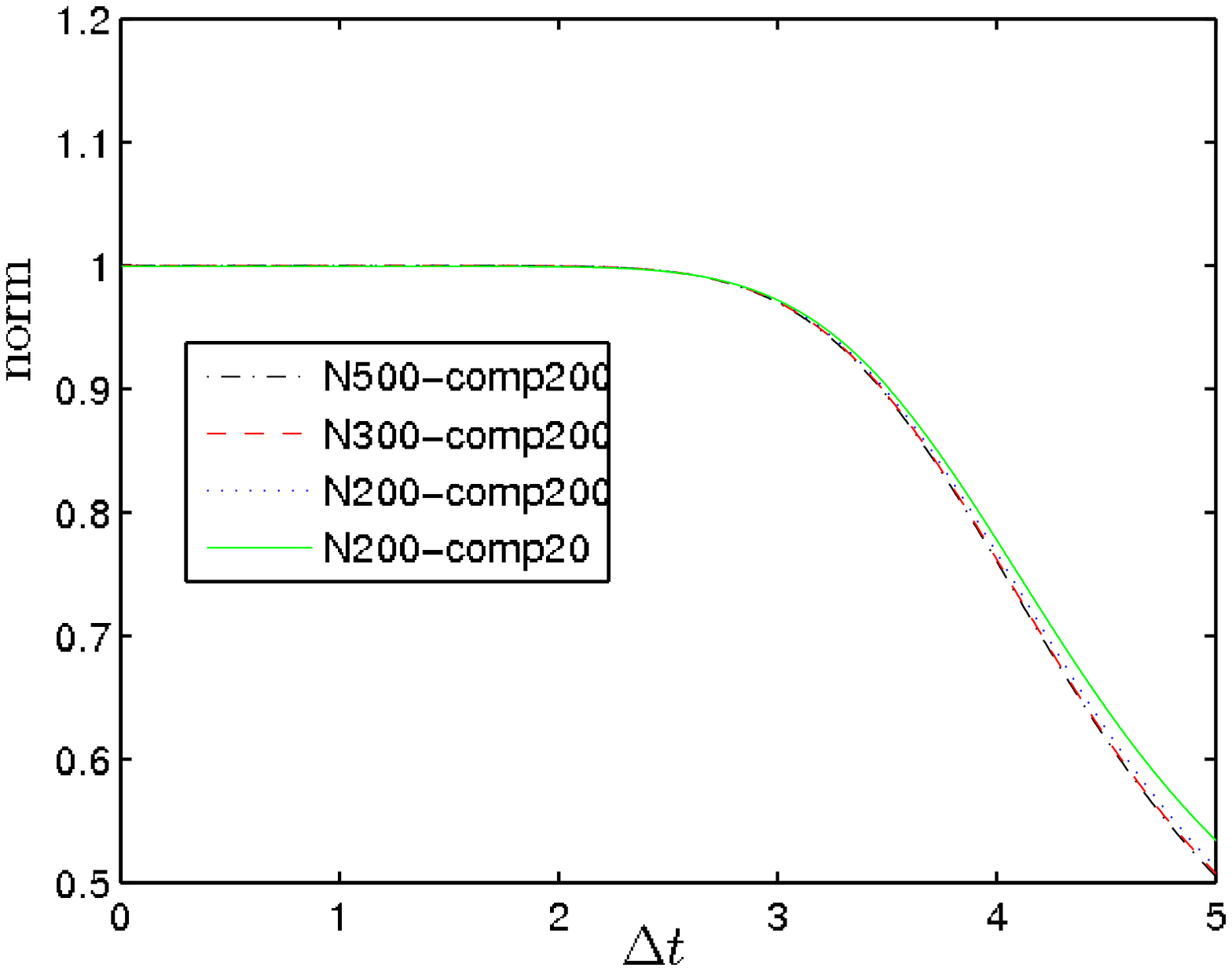}\label{norm}}
\subfigure[Energy versus rescaled time]{\includegraphics[scale=0.26]{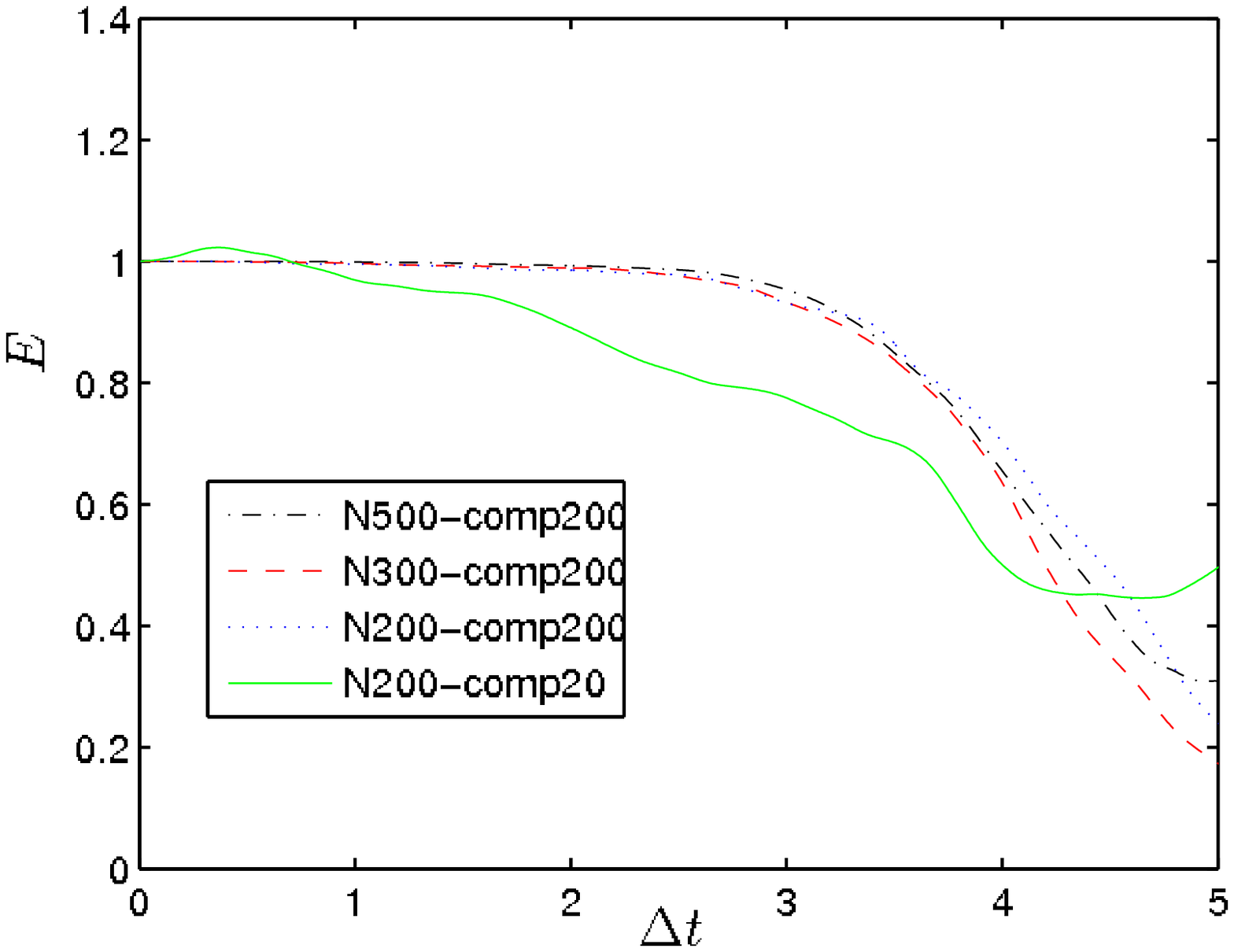}
\label{ene}}
\caption{Norm and expectation value of the energy for MCE results at zero temperature,
for $\hat{H}$ with $\varepsilon=\Delta=g_1=g_2=1$, $M=10$ (with $\omega_m=0.1m$ for $1\le m\le10$) and different values of
$N$ and $comp$.}
\end{center}
\end{figure}

\begin{figure}[t!]
\begin{center}
\subfigure[$\varrho_{11}$ versus rescaled time]{\includegraphics[scale=0.26]{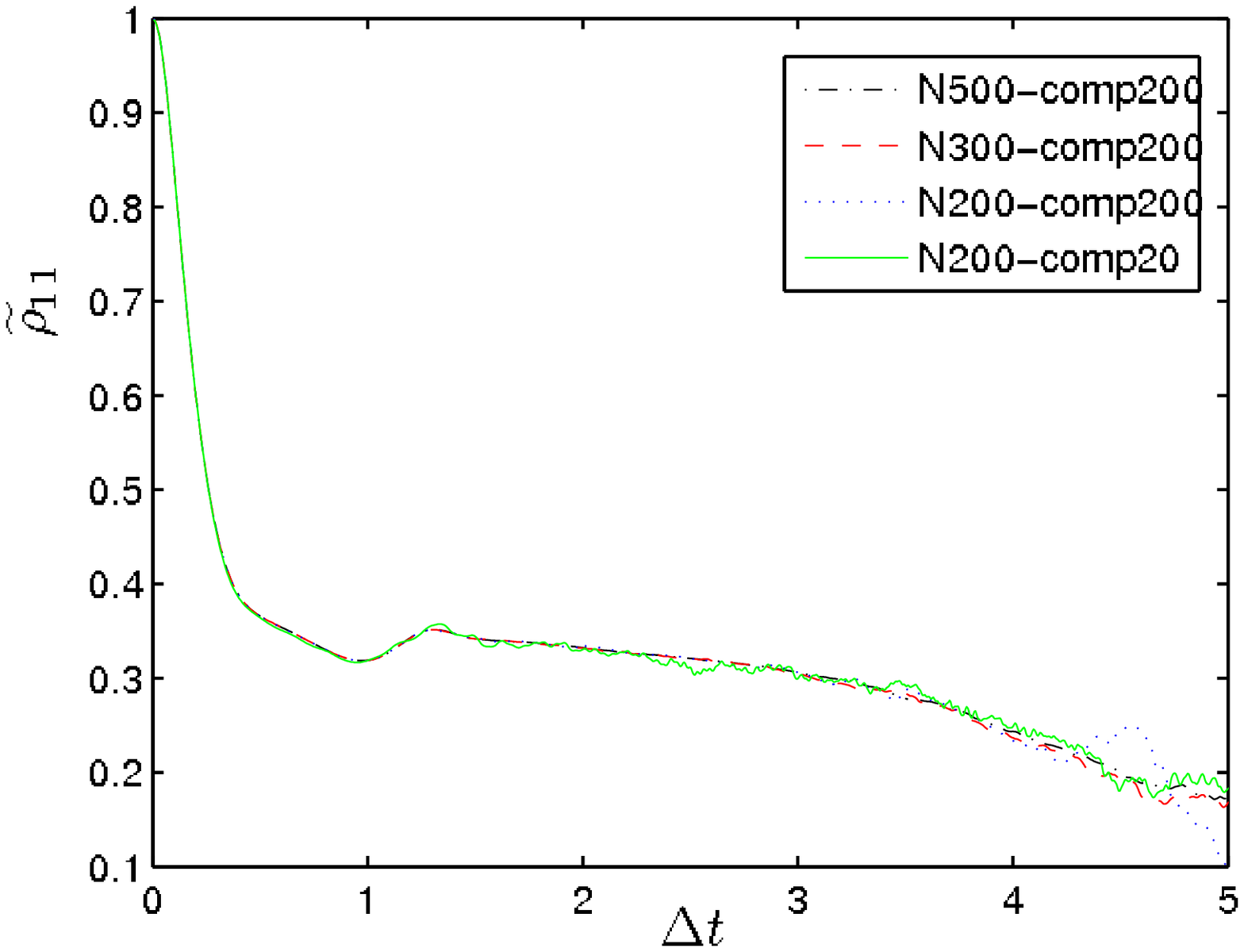}\label{rho11}}
\subfigure[$\varrho_{13}$ versus rescaled time]{\includegraphics[scale=0.26]{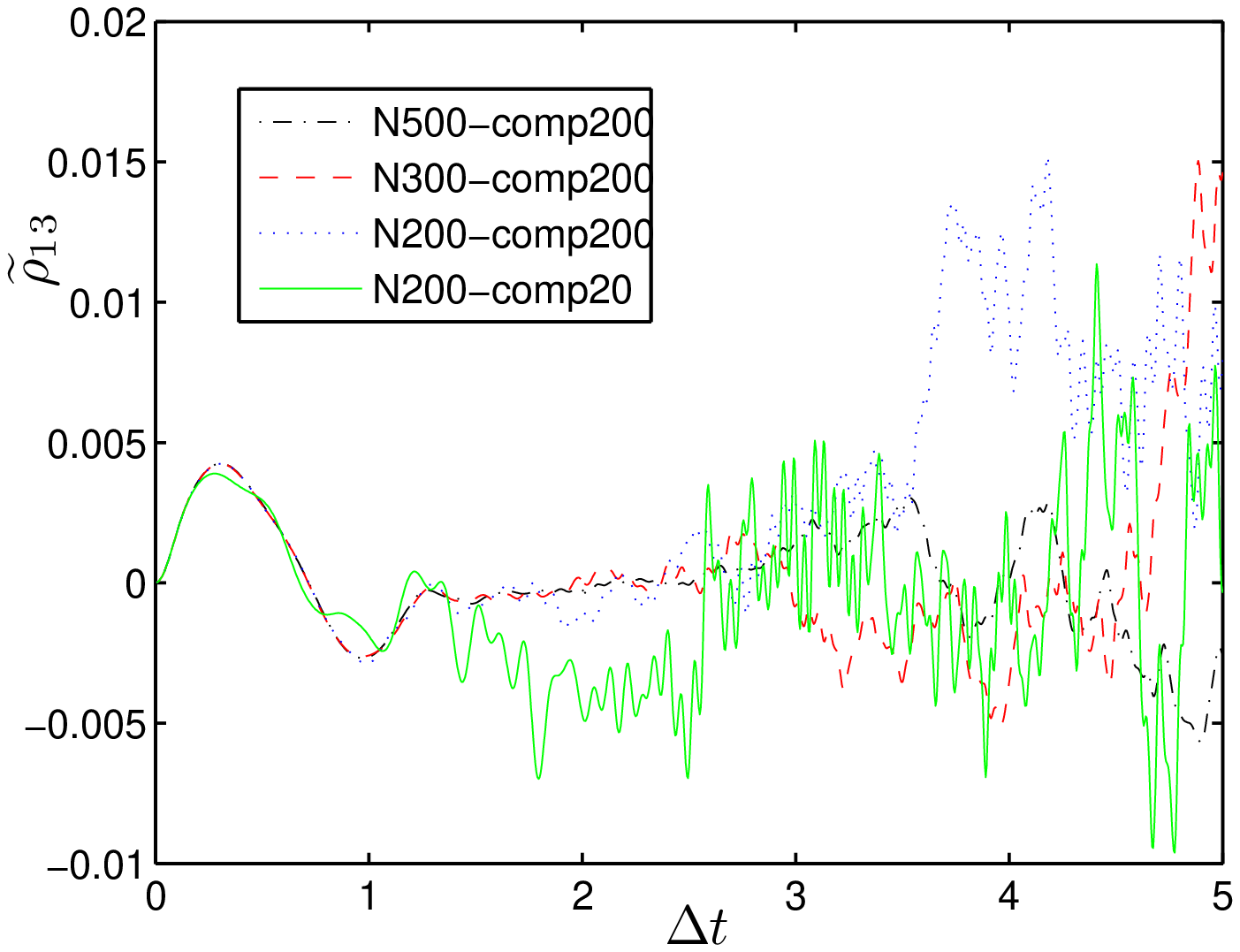}
\label{rho13}}
\caption{Entries of the qubits' density matrix $\varrho_{11}$ and $\varrho_{13}$ for MCE results at zero temperature,
for $\hat{H}$ with
$\varepsilon=\Delta=g_1=g_2=1$, $M=10$ (with $\omega_m=0.1m$ for $1\le m\le10$) and different values of
$N$ and $comp$.}
\end{center}
\end{figure}

\begin{figure*}[t!]
\begin{center}
\subfigure[Choi fidelity]{\includegraphics[scale=0.34]{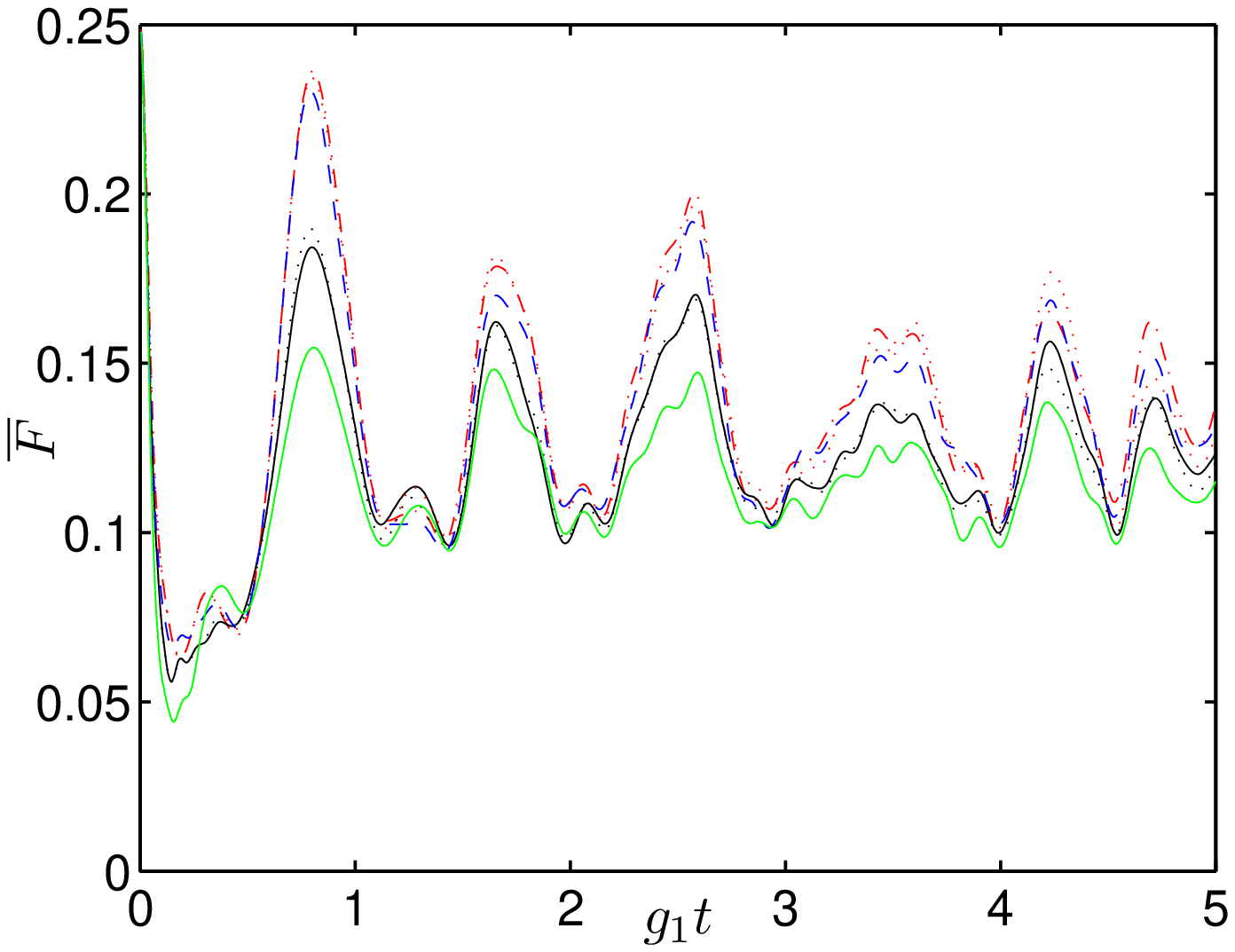}\label{convfid}}
\subfigure[Concurrence for initial state $\ket{\uparrow\uparrow}$]{\includegraphics[scale=0.34]{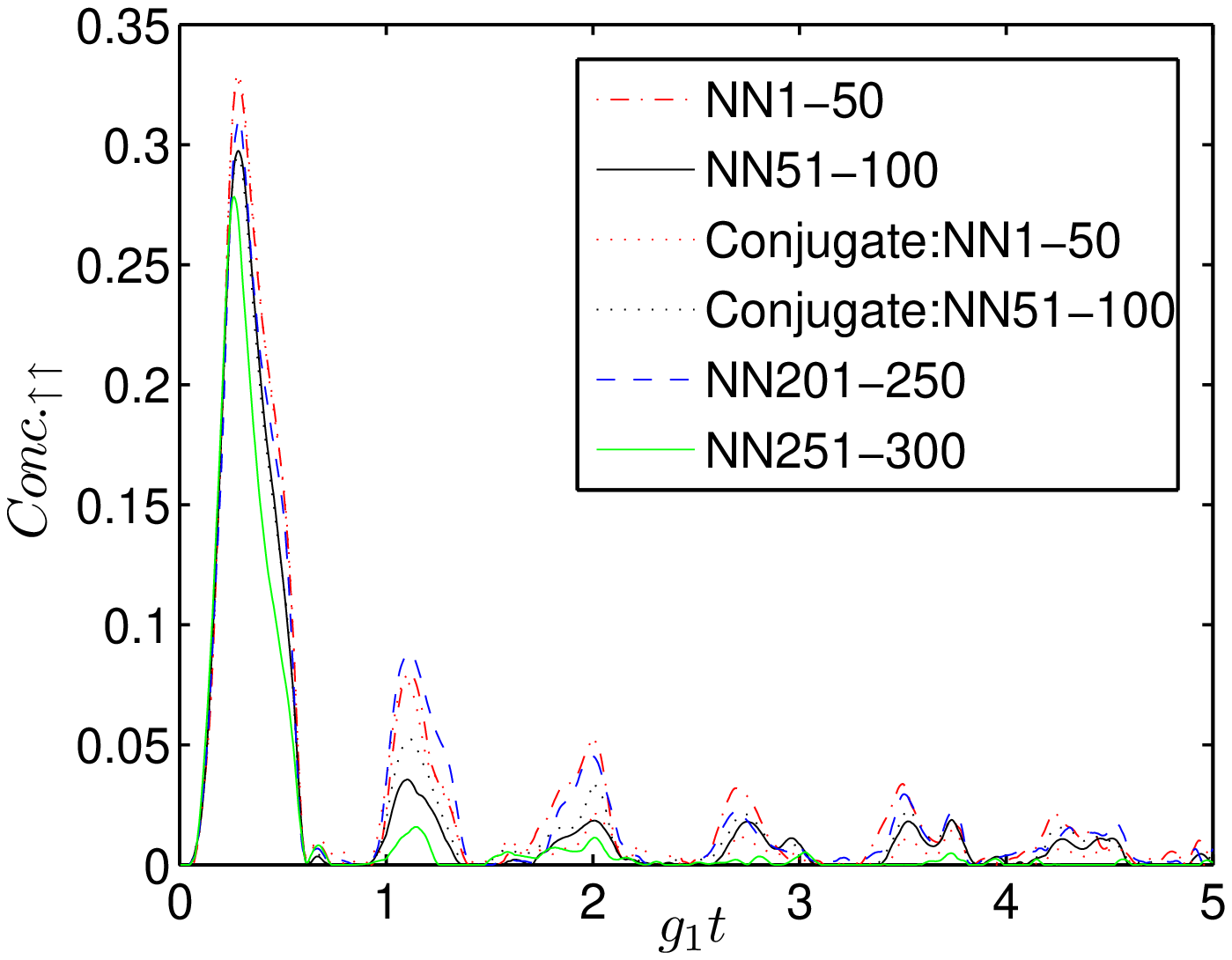}\label{convent1}}
\subfigure[Concurrence for initial state $\ket{\uparrow\downarrow}$]{\includegraphics[scale=0.34]{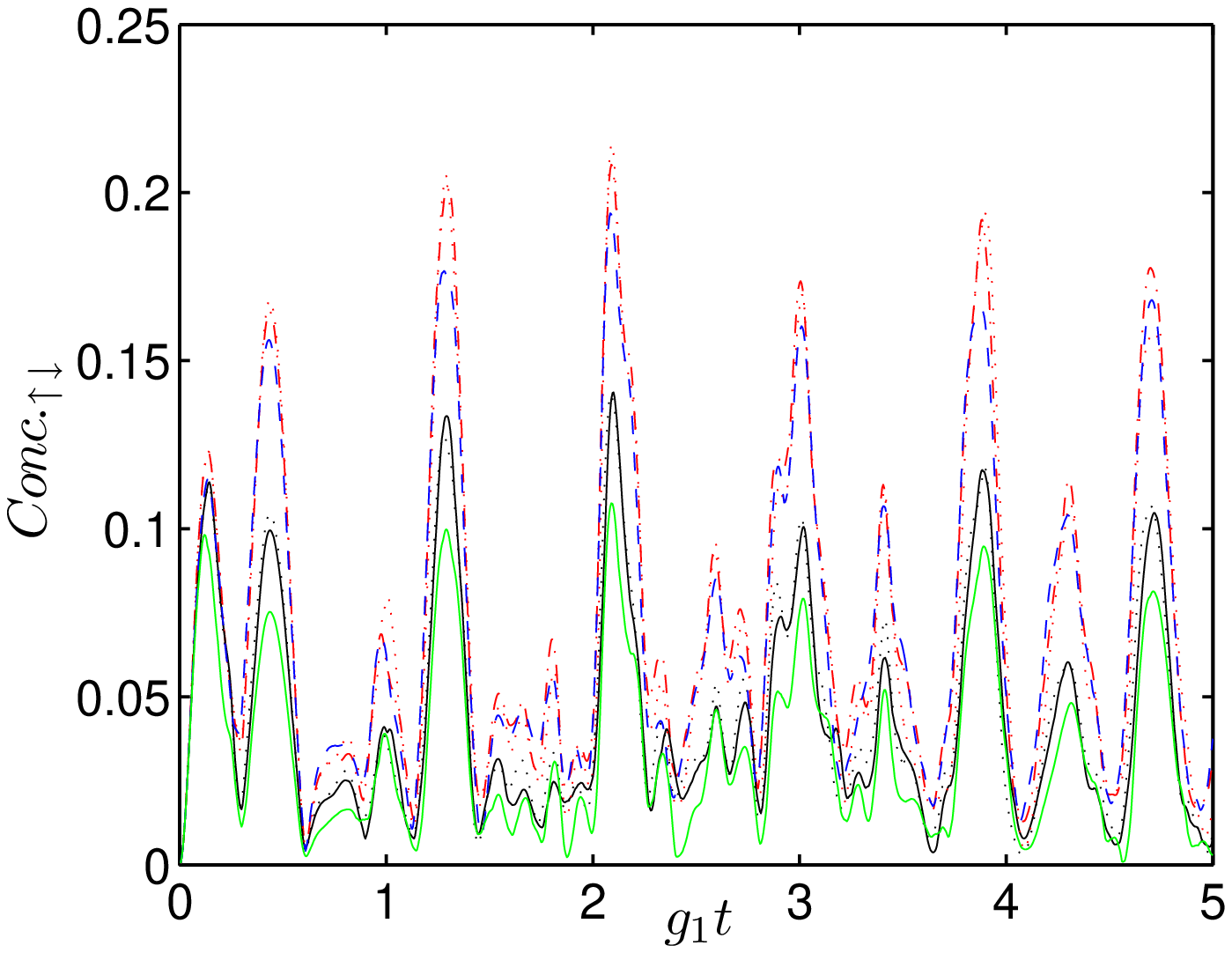}\label{convent2}}
\caption{Choi fidelity and concurrence for two different separable initial states versus rescaled time,
for $\hat{H}_{rw}$ with $\varepsilon=\Delta=0$, $g_1=1$, $g_2=2.1$, $\beta=0.5$ and different values of $N_T$ and $M$.
In all plots $M=10$ with $\omega_m=0.1m$ for $1\le m\le10$. ``Conjugate'' refers to the fact that for those curves the initial
centres of the coherent states $\vec{\alpha}_j$ are in complex conjugate pairs.}
\end{center}
\end{figure*}

\appendix

\section{Ehrenfest dynamics of the coherent states \label{ehren}}
Here, we elaborate on Eq.~(\ref{simpleton}) and derive explicitly the equation of motion of
each complex parameter $\alpha^{(m)}_{j}$.
The approximated Lagrangian ${\mathcal L}_j$ reads
\bea
{\mathcal L}_j &=& \sum_{l,n=1}^{d} c_{l,j}^{*}c_{n,j} \bra{\vec{\alpha}_j,l} \hat{H} \ket{\vec{\alpha}_j,n}
- i \sum_{l=1}^{d} c_{l,j}^{*}\dot{c}_{l,j} \nonumber \\
&& - i \sum_{l=1}^{d} |c_{l,j}|^2 \left( \frac{\dot{\vec{\alpha}}_j \cdot \vec{\alpha}_j^{*}}{2} -
\frac{\dot{\vec{\alpha}}_j^{*} \cdot \vec{\alpha}_j}{2} \right) \; ,
\eea
where $\ket{l,\vec{\alpha}_j}=\ket{l}\otimes\ket{\vec{\alpha}_j}$. Note that
$\bra{\vec{\alpha}_j,l} \hat{H} \ket{\vec{\alpha}_j,n}$ is just a function of the vector $\vec{\alpha}_j$ and the
integers $l$ and $n$, promptly evaluated by normal ordering $\hat{H}$.

The Euler-Lagrange equation for $\alpha_{j}^{(m)}$ then is:
\bea
\frac{\partial{\mathcal L}_j}{\partial \alpha_j^{(m)}} &=& \sum_{l,n=1}^{4} c_{l,j}^{*}c_{n,j} \frac{\partial}{\partial \alpha_{j}^{(m)}}
\bra{\vec{\alpha}_j,l} \hat{H} \ket{\vec{\alpha}_j,n} \nonumber \\
&& + i \sum_{l=1}^{d} |c_{l,j}|^2 \frac{\dot{\alpha}_{j}^{(m)*}}{2} = \\
&&= -i \frac{{\rm d}}{{\rm d}t}\left( \sum_{l=1}^{d} |c_{l,j}|^2 \frac{\alpha^{(m)*}_j}{2} \right) =
\frac{{\rm d}}{{\rm d}t}\frac{\partial{\mathcal L}_j}{\partial \dot{\alpha}_{j}^{(m)}}  \nonumber
\eea
which, by neglecting the time dependence of $c_{l,j}$ (setting $\frac{{\rm d}}{{\rm d}t}\left( \sum_{l=1}^{d} |c_{l,j}|^2\right)=0$),
can be rearranged to obtain
\be
\dot{\alpha}^{(m)*}_j = i \frac{\sum_{l,n=1}^{4} c_{l,j}^{*}c_{n,j} \frac{\partial}{\partial \alpha_{j}^{(m)}}
\bra{\vec{\alpha}_j,l} \hat{H} \ket{\vec{\alpha}_j,n}}{\sum_{l=1}^{d} |c_{l,j}|^2} \, .
\ee
This equation
governs the evolution of the vectors $\vec{\alpha}_j$ in our numerics.

\section{Dynamics of the state vector \label{schroed}}

For the sake of completeness, let us also report the system of dynamical equations
(Schr\"odinger equation on the subspace spanned by the basis grid) for the
parameters $c_{l,j}$, which can be derived by Eq.~(\ref{fullag}) and reads
\bea
\sum_{k=1}^{N}\Bigg[i \Omega_{jk}\dot{c}_{l,k}+
i \Omega_{jk}\left(\vec{\alpha}_j^{\dag}\dot{\vec{\alpha}}_k - \frac{\vec{\alpha}_j^{\dag}\dot{\vec{\alpha}}_j}{2} -
\frac{\dot{\vec{\alpha}}_j^{\dag}\vec{\alpha}_j}{2} \right) c_{l,k} \nonumber \\
- \sum_{h=1}^{d}\bra{\vec{\alpha}_j,l}\hat{H}\ket{h,\vec{\alpha}_k} \Omega_{jk} c_{h,k}
\Bigg] = 0 \; , \nonumber
\eea
where $\Omega_{jk}=\bra{\vec{\alpha}_j}\vec{\alpha}_k\rangle$.
To enhance the stability of the numerical treatment,
the parameters $c_{l,j}$ are actually redefined by multiplication with a smooth phase factor
(essentially a semi-classical action).

\section{Convergence of MCE results}

To give an idea of the quality and range of reliability of our results, we provide here some evidence
of the convergence of our numerics.

Throughout our study, the centres of the initial set of coherent states are distributed in phase space with
a Gaussian distribution with standard deviation $1/comp$. The parameter $comp$ is a free parameter of the study,
which is tuned to optimise convergence.
As indicators of the quality of the numerics we will observe the convergence of specific entries of the
density matrix of the two qubits $\varrho$, as well as the `norm' ${\rm Tr}(\varrho)$ and the expectation
value of the energy ${\rm Tr}(\hat{H}\varrho)$, which are obviously conserved in the exact dynamics.
Notice that our method does not have any in-built routine guaranteeing the conservation of the state vector's norm,
so that ${\rm Tr}(\varrho)$ is a relevant figure of merit to assess its reliability.

Figs.~\ref{normrw} and \ref{enerw} display the norm and expectation value of the energy for a case of
non-number conserving rotating wave Hamiltonian, while in Figs.~\ref{rho11rw} and \ref{rho13rw}
the entries $\varrho_{11}$ and $\varrho_{13}$ are plotted. The reliability of the numerics
over the whole timeframe considered is apparent (for large enough compression parameter $comp$),
in terms of both convergence with increasing
number $N$ of coherent states and of conservation of invariant quantities.
As anticipated, the situation is much more dire for the full Hamiltonian $\hat{H}$.
In this case, Figs.~\ref{norm}, \ref{ene}, \ref{rho11} and \ref{rho13}) show that our numerics are only reliable up to
rescaled times around $2.5$, after which both convergence, and norm and energy conservation are lost,
even at smaller coupling strengths (in that $g_2=1$ rather than $g_2=2.7$ as before).
Finally, we show three examples of convergence of our results at finite temperature (here, $\beta=0.5$)
with respect to the increase in the number of states $N_T$ over which the thermal distribution of Eq.~(\ref{gs})
is sampled (Figs.~\ref{convfid} ,\ref{convent1} and \ref{convent2}).



\begin{references}
\bibitem{lastfootnote} {To be precise, in our instance, since the infinite dimensional Hilbert space is only approximately represented by a basis of $4N$ vectors, such equations are equivalent to the projection of Schr\"odinger equation
on the subspace spanned by our basis.}
\end{references}
\end{document}